\documentclass[fleqn,usenatbib]{mnras}

\usepackage{newtxtext,newtxmath}
\usepackage{multirow}

\usepackage[T1]{fontenc}

\usepackage{amsmath}

\usepackage{graphicx}	
\usepackage{amsmath}		
\usepackage{graphics}
\usepackage{caption,subcaption}
\usepackage{geometry}
\usepackage{tabularx}
\usepackage{fixltx2e}
\usepackage{natbib}
\usepackage{footnote}
\usepackage{tablefootnote}
\usepackage[utf8]{inputenc}
 
\title[Low CI/CO Abundance Ratio]{Low CI/CO Abundance Ratio Revealed by HST UV Spectroscopy of CO-rich Debris Disks}

\author[]{Aoife Brennan$^{1,2}$\thanks{E-mail:brenna29@tcd.ie},
    Luca Matr\`a$^{1}$,
    Sebasti\'an Marino$^{3}$,
    David Wilner$^{4}$,
    Chunhua Qi$^{4}$,
    A. Meredith Hughes,$^{5}$
    \newauthor
    Aki Roberge$^{6}$,
    Antonio S. Hales$^{7}$,
    Seth Redfield$^{5}$
\\
$^{1}$ School of Physics, Trinity College Dublin, Dublin 2, Ireland\\
$^{2}$ European Southern Observatory, Alonso de C$\acute{o}$rdova 3107, Casilla 763 0355, Santiago, Chile\\
$^{3}$ Department of Physics and Astronomy, University of Exeter, Stocker Road, Exeter, EX4 4QL, UK\\
$^{4}$ Center for Astrophysics | Harvard \& Smithsonian, 60 Garden St., Cambridge, MA 02138, USA\\
$^{5}$ Department of Astronomy, Van Vleck Observatory, Wesleyan University, 96 Foss Hill Dr., Middletown, CT 06459, USA\\
$^{6}$ Exoplanets and Stellar Astrophysics Lab, NASA Goddard Space Flight Center, Greenbelt, MD 20771, USA\\
$^{7}$ National Radio Astronomy Observatory, 520 Edgemont Road, Charlottesville, VA 22903-2475, USA \\
}

\date{Accepted 2024 May 14. Received 2024 May 14; in original form 2023 September 30}

\pubyear{2024}

\begin{document}
\label{firstpage}
\pagerange{\pageref{firstpage}--\pageref{lastpage}}
\maketitle

\begin{abstract}
The origin and evolution of CO gas in debris disks has been debated since its initial detection. The gas could have a primordial origin, as a remnant of the protoplanetary disk or a secondary exocometary origin. This paper investigates the origin of gas in two debris disks, HD110058 and HD131488, using HST observations of CI and CO, which play critical roles in the gas evolution. We fitted several electronic transitions of CI and CO rovibronic bands to derive column densities and temperatures for each system, revealing high CO column densities ($\sim$ 3-4 orders of magnitude higher than $\beta$ Pictoris), and low CI/CO ratios in both. Using the \textsc{exogas} model, we simulated the radial evolution of the gas in the debris disk assuming a secondary gas origin. We explored a wide range of CO exocometary release rates and $\alpha$ viscosities, which are the key parameters of the model. Additionally, we incorporated photodissociation due to stellar UV to the \textsc{exogas} model and found that it is negligible for typical CO-rich disks and host stars, even at a few au due to the high radial optical depths in the EUV. We find that the current steady-state secondary release model cannot simultaneously reproduce the CO and CI HST-derived column densities, as it predicts larger CI/CO ratios than observed. Our direct UV measurement of low CI/CO ratios agrees with results derived from recent ALMA findings and may point to vertical layering of CI, additional CI removal, CO shielding processes, or different gas origin scenarios.

\end{abstract}


\section{Introduction}

Protoplanetary disks rich in gas and dust evolve into debris disks similar to our solar system's Kuiper belt. Traditionally, debris disks were thought to be depleted of dust and completely exhausted of gas. However, thanks largely to the Atacama Large Millimeter/submillimeter Array (ALMA), gas has now been detected in $\sim$ 20 debris disks. The majority of these detections are CO emission lines, in which the cold gas is co-located with the debris disks at tens of au \citep[e.g.][]{Marino(2016), Matra(2019)}. Additionally, debris disk gas has been detected through UV and optical absorption lines against the stellar background in edge-on systems \citep[e.g.][]{S.Redfield(2007), Iglesias(2018), rebollido(2018)}. These lines are typically observed at the stellar velocity. However, variable red or blue-shifted absorption is also observed in some systems and is indicative of falling evaporating bodies at a few stellar radii \citep[e.g.][]{Ferlet(1987), Beust(1990), Crawford(1994), Kiefer(2014)}. 

The cold gas in debris disks could be a long-lived remnant of the protoplanetary disk (referred to as primordial) and predominantly composed of H$_2$ \citep{Nakatani(2021)}. Conversely, the gas could be of secondary origin, released by exocomets within the debris disk's collisional cascade, and therefore have a cometary (H$_2$-poor) composition. Several disks with low CO gas levels are considered to be of secondary origin. This is due to the required constant replenishment of their low gas levels, as the interstellar radiation field (ISRF) alone should cause unshielded CO to photodissociate within $\sim$ 130 years \citep[][]{Heays(2017)}. Due to the low CO levels, neither CO nor unobservable H$_2$ (even with interstellar-medium like H$_2$/CO ratios) could shield CO over the system age. Therefore, the observed low CO gas levels cannot be primordial and must be continuously replenished; this is the case for $\beta$ Pictoris, HD 181327, Fomalhaut, and $\eta$ Corvi \citep[e.g.][]{Marino(2016), Matra(2017), Marino(2017)}, among others.

On the other hand, some debris disks exhibit large CO mass levels that are comparable to those found in young protoplanetary disks \citep[][]{Moor(2017)}. As the dust levels in debris disks are not sufficient to provide shielding, molecular hydrogen and CO were initially proposed as shielding agents in a primordial gas scenario. This led to these disks being termed hybrid, due to the coexistence of primordial gas with second-generation dust \citep[][]{Kospal(2013)}. In this hybrid disk scenario, primordial CO is shielded by large amounts of unseen H$_2$, allowing it to survive since the protoplanetary phase of evolution. 

\citet[][]{Q.Kral(2019)} proposed a secondary origin for the gas in debris disks with high observed CO levels. In this scenario, CO gas is released in the disk due to colliding exocomets and then photodissociates into C and O. This largely atomic gas disk radially spreads viscously over time, eventually leading to accretion onto the star. The key component of this model is the shielding of CO by photoionization of neutral carbon (CI), which allows high CO levels to accumulate reproducing the high masses observed. The shielding is most effective when the CI column density is highest, which occurs if viscosities are low (slow radial spreading) or if the CO input rate is high \citep[in turn producing CI faster, e.g.][]{Q.Kral(2019),S.Marino(2020)}. Therefore, the introduction of CI shielding could allow for a secondary origin of gas to explain high-mass as well as low-mass CO disks. \citet[][]{Cataldi(2023)} found that this secondary model with CI shielding overpredicts the CI masses derived from ALMA emission data of a number of CO-rich disks. This suggests that alternative shielding mechanisms, origin scenarios, and/or effective CI removal other than through viscous accretion may be at play.

As well as the gas viscosity and input rate from the belt, the vertical distribution of CI strongly affects the shielding vertically \citep[][]{S.Marino(2022)}. For example, if the gas is vertically layered with CI in a narrow layer above and below the CO midplane, the CI gas can most effectively provide shielding vertically from the ISRF. Conversely, when CO and CI are mixed vertically throughout the midplane the effectiveness of CI shielding is decreased. Whether we are in a layered or mixed scenario depends on the vertical diffusion, which can be tested with future observations. 

Our work here focuses on constraining models of gas in debris disks by accurately measuring the column density, temperature, and relative abundance of CO and CI in two debris disks (HD110058 and HD131488) using absorption spectroscopy against the host star's background with the \textit{Hubble} Space Telescope (HST). Due to the high CO levels observed by ALMA and edge-on nature, these two systems are optimal HST targets to investigate the gas origin and evolution in debris disks \citep[e.g.][]{Moor(2017), A.Hales(2022)}. 

HD110058 is a 17 Myr-old A-type star that is part of the Lower Centaurus Crux of the Scorpius-Centaurus association, with an associated distance of $129.9_{-1.2}^{+1.3}$ pc \citep[][]{Gaia(2018), Luhman(2020)}. The disk is rich in both dust and CO gas, with measured masses of $0.080_{-0.003}^{+0.002}$ M$_\oplus$ and $0.069_{-0.057}^{+3.56}$ M$_\oplus$ \citep[][]{A.Hales(2022)}. The dust extends to 70 au, peaking at $\sim$ 30 au and has the smallest inner edge (18 au) observed by ALMA thus far. The CO gas distribution is more compact than the dust but with a peak radius consistent with the dust’s peak radius. 

HD131488 is a 16 Myr-old A-type star located in the Upper Centaurus Lupus subgroup of the Scorpius-Centaurus association, with an associated distance of $151.4_{-0.8}^{+0.6}$ pc \citep[][]{Gaia(2018), Luhman(2020)}. This disk also has high levels of dust and CO gas detected by ALMA, with a dust mass of 0.32 M$_\oplus$ and a gas mass of $0.089\pm0.015$ M$_\oplus$ \citep{Moor(2017)}. The dust disk radius is ($168\pm6$ au), with a ring width (FWHM) of $44\pm11$ au. The CO gas disk has an inner radius of 35 au and an outer radius of 140 au \citep[][]{Smirnov-Pinchukov(2022)}. 

In §2 we describe the HST observations. In §3 we present the observational results and describe our absorption model, which is used to simulate absorption lines that are then fitted to the observations. In §4, we discuss the latest secondary gas evolution model and present our additions to the model. In §5, we interpret our measured column densities for the two systems in the context of the updated secondary gas model. Finally, in section §6, we discuss our results and investigate possible gas origins for both disks.

\section{HST Observations}

We aim to model CO and CI absorption lines observed with HST across a wide wavelength range of $\sim$ 1220 to 3050 \AA\ over 24 orbits spanning between July 6 and July 26, 2020. The Space Telescope Imaging Spectrograph (STIS) and the Cosmic Origins Spectrograph (COS) were both used to cover wavelengths of atomic and molecular transitions of interest in this range. COS was selected for use below $\sim$ 1500 \AA\ due to its increased sensitivity, while STIS was chosen for use beyond this wavelength for its higher resolving power (R $\sim$ 114000\footnote{STIS resolution from \url{https://hst-docs.stsci.edu/stisihb/chapter-3-stis-capabilities-design-operations-and-observations/3-1-instrument-capabilities}} compared to R $\sim$ 14000\footnote{COS resolution from \url{https://hst-docs.stsci.edu/cosihb/chapter-2-proposal-and-program-considerations/2-1-choosing-between-cos-and-stis}} for COS).

\begin{table*}
\caption{Table of observations that are processed in this paper.}
\begin{tabular}{|c|c|c|c|c|c|c|}
\hline
System                    & Instrument            & Detector & Grating & Slit (arcsec)             & Central wavelength (\AA) & Exposure time (sec) \\ \hline
\multirow{3}{*}{HD110058} & \multirow{2}{*}{STIS} & FUV-MAMA & E140H   & \multirow{2}{*}{0.2 $\times$ 0.09} & 1562                                   & 7701                \\ \cline{3-4} \cline{6-7} 
                          &                       & NUV-MAMA & E230H   &                           & 2912                                   & 1226                \\ \cline{2-7} 
                          & COS                   & FUV      & G130M   & Slitless                  & 1222                                   & 7588                \\ \hline
\multirow{3}{*}{HD131488} & \multirow{2}{*}{STIS} & FUV-MAMA & E140H   & \multirow{2}{*}{0.2 $\times$ 0.09} & 1562                                   & 2906                \\ \cline{3-4} \cline{6-7} 
                          &                       & NUV-MAMA & E230H   &                           & 2912                                   & 1226                \\ \cline{2-7} 
                          & COS                   & FUV      & G130M   & Slitless                  & 1222                                   & 1593                \\ \hline
\end{tabular}
\label{table:observations}
\end{table*}

Observation details are provided in Table \ref{table:observations}, including the instrument, grating, central wavelength, slit, science target system, and exposure time. The data used in this paper was reduced and calibrated through the HST pipeline using the Space Telescope Science Institute (STScI) IRAF package CALSTIS\footnote{Information on the pipeline can be found at \url{https://hst-docs.stsci.edu/stisdhb/chapter-3-stis-calibration/3-1-pipeline-processing-overview}}  (version 3.4.2).

\section{Results \& Modelling}
\subsection{Results}

We focus on the absorption lines observed by STIS, as the star is brighter at longer wavelengths, which allows for a higher signal-to-noise ratio (SNR) and greater sensitivity to faint narrow lines. Moreover, the higher spectral resolution of STIS allowed us to effectively distinguish between neighbouring transitions within CO bands. The detected CI transitions and CO bands in the COS data were not fitted as they were either saturated and/or less constraining at the much lower spectral resolution of COS.

To identify CI absorption lines in our study, we manually searched for lines using the National Institute for Standards and Technology (NIST) database. We selected lines that were either detected but not saturated or near the detection limit. Ultimately, we selected a CI triplet at 1613.38 \AA\ and a CI doublet at 2965.71 \AA\ (Table \ref{table:lines}). The CI triplet at 1613.38 \AA\ (referred to as the "CI triplet" in this paper, see Fig. \ref{fig:Cspectra}) was found to be moderately optically thick for both systems. In contrast, the CI doublet at 2965.71 \AA\ (referred to as the "CI doublet" in this paper) was only detected for HD110058 (see Fig. \ref{fig:Cspectra}).

We detected a large number of CO bands, including the main CO absorption bands $A^{1}\scriptstyle\prod(v) - X'\scriptstyle\sum ^{+}(0)$ $\sim$ 1300 to 1600 \AA\ \citep[][]{Morton(1994)}. However, these bands are highly saturated which caused significant broadening of the line to the extent that the continuum cannot be reliably determined. We chose to fit all detected CO intersystem bands from the MOLAT database\footnote{Available at \url{https://molat.obspm.fr/index.php?page=pages/menuSpectreMol.php}} with oscillator strengths between $1 \times 10^{-7}$ and $1 \times 10^{-5}$ for HD110058 and $1 \times 10^{-5}$ and $2.5 \times 10^{-5}$ for HD131488. These oscillator strengths cover bands from just below the detection limit, to marginally detected bands, up to moderately optically thick bands, ensuring we sample a broad range of line depths while ensuring we can still reliably separate individual rovibronic lines and define the continuum.  In Table \ref{table:lines}, the fitted CO bands are listed for each system. We only excluded detected bands within these oscillator strength ranges that were contaminated by strong, broad stellar absorption lines, to prevent bias in our column density and temperature values.

We detected intrinsically weaker bands with lower oscillator strengths (Table \ref{table:lines}) for HD110058 compared to HD131488, due to higher column densities (see section 3.4). In total, we fitted twelve CO bands (Fig. \ref{fig:COspectra} and \ref{fig:HD110_CO}) for HD110058, and five CO bands (Fig. \ref{fig:COspectra} and \ref{fig:HD131_CO}) for HD131488. We normalised the spectra by the continuum for each CO band or CI triplet/doublet by fitting a polynomial of suitable order to line-free regions of the spectrum. The optimal polynomial order for each band was determined visually by selecting the order that produced a flat continuum-normalised spectrum while avoiding overfitting and introducing noticeable fluctuations. We consistently opt for the lowest-order polynomial that provides a good fit. The wavelengths, energy levels with their degeneracies, and oscillator strengths for both the CI doublet and triplet were obtained from the NIST\footnote{Degeneracies, wavelengths, and energy levels available at \url{https://www.nist.gov/pml/atomic-spectra-database}} database, while the energy levels, degeneracies, wavelengths, and oscillator strengths for all CO bands were sourced from the MOLAT\footnote{Energy levels, degeneracies, wavelengths, and oscillator strengths available at \url{https://molat.obspm.fr/index.php?page=pages/menuSpectreMol.php}} database.

\begin{table*}
\caption{Table of the absorption lines and bands that were modeled in this paper. The CO bands listed are composed of several transitions, while the CI lines listed are single transitions. For the CO bands, the wavelengths and oscillator strengths listed correspond to the wavelengths and oscillator strength of the first transition in the band, where the oscillator strength is calculated at 300K \citep[][]{Rostas(2000)}.}
\begin{tabular}{|c|l|l|l|l|}
\hline
Species              & \multicolumn{1}{c|}{Band/Line} & \multicolumn{1}{c|}{Oscillator Strength} & \multicolumn{1}{c|}{Wavelength (\AA)} & \multicolumn{1}{c|}{Fitted System}
 \\ \hline
\multirow{13}{*}{CO}          & \multicolumn{1}{c|}{$d^{3}D(7) - X^{1}S^{+}(0)$}    & \multicolumn{1}{c|}{2.24 $\times$ 10$^{-5}$}     & \multicolumn{1}{c|}{1464.12}   & \multicolumn{1}{c|}{Both} \\
                              & \multicolumn{1}{c|}{$e^{3}S^{-}(4) - X^{1}S^{+}(0)$}   & \multicolumn{1}{c|}{4.68 $\times$ 10$^{-5}$}    & \multicolumn{1}{c|}{1470.97}   & \multicolumn{1}{c|}{Both} \\
                              & \multicolumn{1}{c|}{$a^{'3}S^{+}(11) - X^{1}S^{+}(0)$}  & \multicolumn{1}{c|}{2.25 $\times$ 10$^{-5}$}       & \multicolumn{1}{c|}{1480.81}   & \multicolumn{1}{c|} {HD131488} \\
                              & \multicolumn{1}{c|}{$d^{3}D(6) - X^{1}S^{+}(0)$}    & \multicolumn{1}{c|}{4.41 $\times$ 10$^{-6}$}     & \multicolumn{1}{c|}{1486.60}   & \multicolumn{1}{c|} {HD110058} \\
                              & \multicolumn{1}{c|}{$e^{3}S^{-}(3) - X^{1}S^{+}(0)$}    & \multicolumn{1}{c|}{2.14 $\times$ 10$^{-6}$}     & \multicolumn{1}{c|}{1493.76}   & \multicolumn{1}{c|}  {HD110058} \\
                              & \multicolumn{1}{c|}{$a^{'3}S^{+}(10) - X^{1}S^{+}(0)$}& \multicolumn{1}{c|}{1.15 $\times$ 10$^{-5}$}         & \multicolumn{1}{c|}{1503.25}   & \multicolumn{1}{c|} {HD110058} \\
                              & \multicolumn{1}{c|}{$e^{3}S^{-}(2) - X^{1}S^{+}(0)$} & \multicolumn{1}{c|}{2.11 $\times$ 10$^{-5}$}        & \multicolumn{1}{c|}{1517.67}  & \multicolumn{1}{c|}{Both}     \\
                              & \multicolumn{1}{c|}{$d^{3}D(4) - X^{1}S^{+}(0)$}   & \multicolumn{1}{c|}{2.23 $\times$ 10$^{-5}$}      & \multicolumn{1}{c|}{1535.09}   & \multicolumn{1}{c|}{Both}     \\
                              & \multicolumn{1}{c|}{$a^{'3}S^{+}(8) - X^{1}S^{+}(0)$}    & \multicolumn{1}{c|}{1.69 $\times$ 10$^{-6}$}     & \multicolumn{1}{c|}{1551.62}   & \multicolumn{1}{c|} {HD110058} \\
                              & \multicolumn{1}{c|}{$a^{'3}S^{+}(7) - X^{1}S^{+}(0)$}  & \multicolumn{1}{c|}{5.90 $\times$ 10$^{-7}$}       & \multicolumn{1}{c|}{1577.67}   & \multicolumn{1}{c|} {HD110058} \\
                              & \multicolumn{1}{c|}{$d^{3}D(2) - X^{1}S^{+}(0)$} & \multicolumn{1}{c|}{2.78 $\times$ 10$^{-7}$}       & \multicolumn{1}{c|}{1588.64}   & \multicolumn{1}{c|} {HD110058} \\
                              & \multicolumn{1}{c|}{$a^{'3}S^{+}(6) - X^{1}S^{+}(0)$} & \multicolumn{1}{c|}{2.76 $\times$ 10$^{-7}$}& \multicolumn{1}{c|}{1605.17}   & \multicolumn{1}{c|} {HD110058} \\ 
                              & \multicolumn{1}{c|}{$a^{'3}S^{+}(5) - X^{1}S^{+}(0)$} & \multicolumn{1}{c|}{1.23 $\times$ 10$^{-7}$}& \multicolumn{1}{c|}{1634.19 }   & \multicolumn{1}{c|} {HD110058} \\ \hline
\multirow{5}{*}{CI}           & \multicolumn{1}{c|}{$^{3}\textrm{P}_{0} - ^{1}\textrm{P}_{1}$} & \multicolumn{1}{c|}{4.30 $\times$ 10$^{-5}$}    & \multicolumn{1}{c|}{1613.38}   & \multicolumn{1}{c|}{Both}     \\
                              & \multicolumn{1}{c|}{$^{3}\textrm{P}_{1} - ^{1}\textrm{P}_{1}$} & \multicolumn{1}{c|}{1.09 $\times$ 10$^{-5}$}       & \multicolumn{1}{c|}{1613.80}   & \multicolumn{1}{c|}{Both}     \\
                              & \multicolumn{1}{c|}{$^{3}\textrm{P}_{2} - ^{1}\textrm{P}_{1}$}& \multicolumn{1}{c|}{6.90$\times$ 10$^{-6}$}       & \multicolumn{1}{c|}{1614.52}   & \multicolumn{1}{c|}{Both}     \\
                              & \multicolumn{1}{c|}{$^{5}\textrm{S}^{\circ } - ^{3}\textrm{P}_{1}$}& \multicolumn{1}{c|}{1.90 $\times$ 10$^{-8}$ }         & \multicolumn{1}{c|}{2965.70}    & \multicolumn{1}{c|}{Both}    \\
                              & \multicolumn{1}{c|}{$^{5}\textrm{S}^{\circ } - ^{3}\textrm{P}_{2}$}& \multicolumn{1}{c|}{2.80 $\times$ 10$^{-8}$}       & \multicolumn{1}{c|}{2968.09}  & \multicolumn{1}{c|}{Both}     \\ \hline
\end{tabular}
\label{table:lines}
\end{table*}

\subsection{Modelling}

We used a simple radiative transfer model to model the absorption lines from CO and CI in the edge-on disks along the line of sight to the star. Our model assumes that the gas column can be characterized by four parameters: the gas column density, its average line-of-sight kinetic temperature, its average line-of-sight excitation temperature and its radial velocity\footnote{For the 1517.67 \AA\ band the best fit radial velocity lies outside the associated error in velocity of 1 km s$^{-1}$ \citep[][]{Sahu(1999)} for STIS. No other band displayed this issue, we attribute this to an error in the wavelength calibration for this order. This shift in radial velocity for the 1517.67 \AA\ band was also detected in both systems, hence the radial velocity for 1517.67 \AA\ was fitted as an additional free parameter for both systems. For HD110058, the 1517.67 \AA\ band has a radial velocity of $10.50\pm0.07$ km s$^{-1}$, for HD131488 the radial velocity for the 1517.67 \AA\ band is $3.30\pm0.09$ km s$^{-1}$.}, which is expected to equal that of the star in the absence of radial gas motion such as a wind \citep[][]{Youngblood(2021), Kral(2022)}. We choose, to fit both the kinetic and excitation temperatures as free parameters as Local thermodynamic equilibrium (LTE) is not necessarily applicable to debris disks \citep[][]{A.Roberge(2000), Troutman(2011), L.Matra(2015)}. 

Since absorption depends on the number of molecules/atoms in the lower energy level of each transition, we focused on absorption lines originating from energy levels with energies corresponding to below a few thousand K. This is because we expect that higher energy levels are not significantly populated in the cold environments of our gas-rich debris disks at tens of au. We modelled the absorption lines using Voigt line profiles (see Appendix A for details). We then generated simulated absorption lines for each modelled transition using a wavelength grid finer than the STIS pixel size. 

The generated spectral models were then convolved with the STIS Line spread function (LSF)\footnote{Available at \url{https://www.stsci.edu/hst/instrumentation/stis/performance/spectral-resolution}}. The LSF describes the response of the STIS instrument for each given grating and detector (FUV-MAMA, E140H, 0.2" $\times$ 0.09" for the CO bands and CI triplet, NUV-MAMA, E230H, 0.2" $\times$ 0.09" for the CI doublet). After convolution, we binned our model onto the same wavelength grid as our data. 

We fitted absorption line models for CO and CI separately. Within each fit for CO and CI, all selected lines/bands were simultaneously modeled using a Bayesian Markov Chain Monte Carlo (MCMC) approach. To sample the posterior probability distribution of the model parameters, we used an affine-invariant ensemble sampler implemented through the emcee package \citep[][]{Goodman(2010),Foreman-Mackey(2013)}. We employed uniform priors with well-defined boundaries, ensuring they were sufficiently distanced from the best-fit values. For all fits excluding the CI fits for HD131488 we ran the MCMC with 32 walkers for 2000 steps discarding the first 500. For the CI fits for HD131488 we ran the MCMC with 32 walkers for 5000 steps discarding the first 2000 due to a slower convergence. The posterior distributions obtained for both CI and CO fits for both systems  are given in Fig. \ref{fig:HD110_C_corner}, \ref{fig:HD110_CO_corner}, \ref{fig:HD131_C_corner} and \ref{fig:HD131_CO_corner}. The best-fit values for each species and system are reported in Table \ref{table:best_fits} as the $50\pm34$ percentile range of the probability distribution of each parameter, marginalised over all others.

\subsection{CI}

\begin{figure*}
\includegraphics[width=1\textwidth]{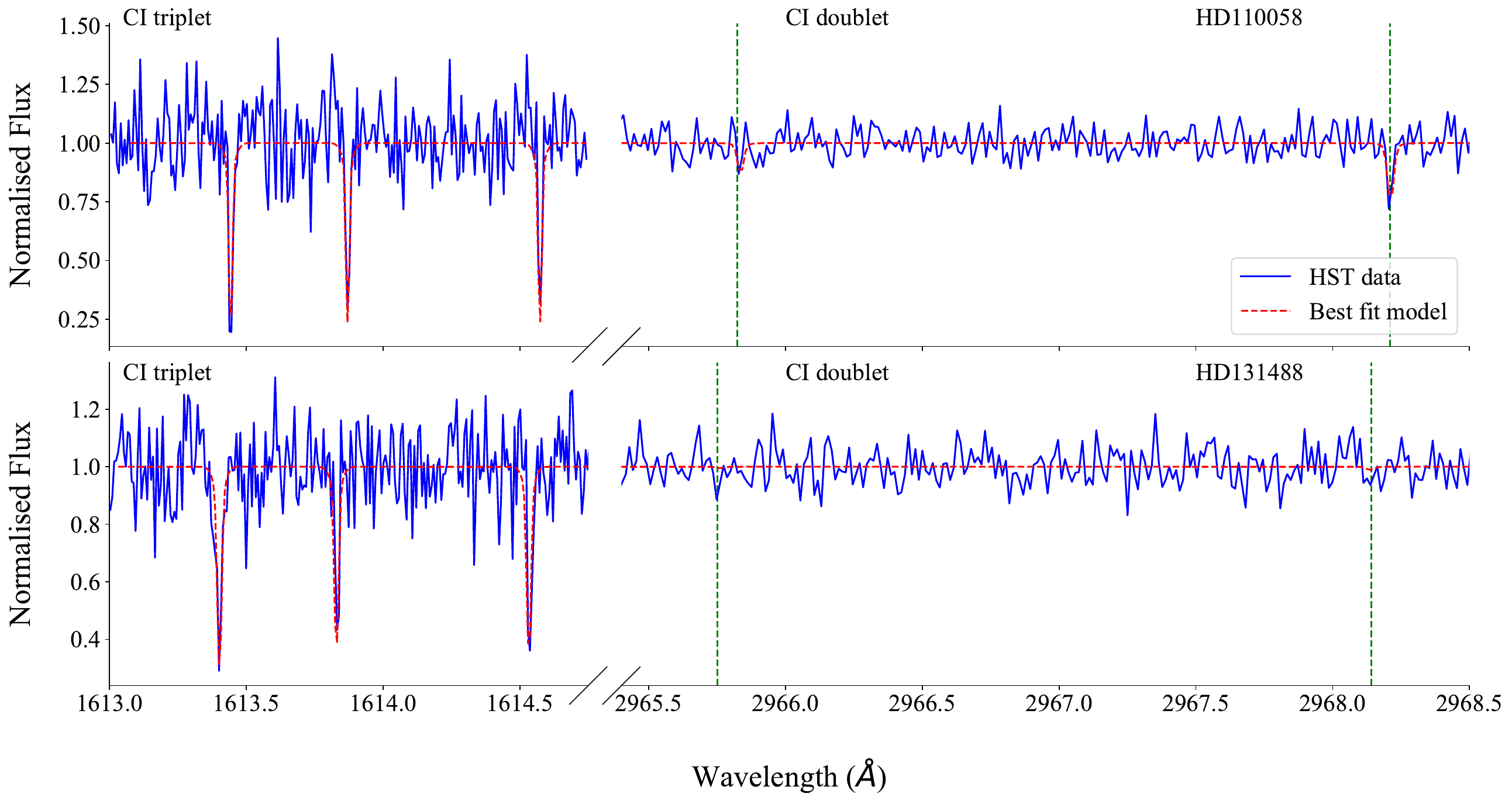}
\caption{The HST continuum-normalised data (blue line) around each transition was fitted with a sixth-order polynomial, and the best-fit model (red line) was overlaid. The left panel shows the CI triplet with fine structure lines detected at 1613.38 \AA, 1613.80 \AA, and 1614.51 \AA, for both HD110058 (top) and HD131488 (bottom). In the right panel, the CI doublet with fine structure lines at 2965.71 \AA, and 2968.09 \AA, is plotted for both HD110058 (top) and HD131488 (bottom). Vertical dashed green lines indicate the position of the doublet for both systems.}
\label{fig:Cspectra}
\end{figure*}

Fig. \ref{fig:Cspectra} shows the continuum-normalised CI triplet and doublet for both systems, plotted alongside the best-fit models generated using the values reported in Table \ref{table:best_fits}. For comparison, Fig. \ref{fig:ComparingDisks} illustrates the CI column densities of all observed edge-on debris disks with HST. Comparing our values with those of $\beta$ Pictoris, we find that the CI column densities in HD110058 are $\sim$ 260 times higher and $\sim$ 9 times higher for HD131488, respectively \citep[][]{A.Roberge(2000)}. Given the higher CI column densities compared to $\beta$ Pictoris, it is possible that CI could act as a braking agent \citep{Roberge(2006), Fernadez(2006)} explaining the presence of the stable Ca II and Na I absorption in both disks \citep{A.Hales(2017), I.Rebollido(2018)}.

It is important to note the degeneracy between column density and kinetic temperature for HD131488 (Fig. \ref{fig:HD131_C_corner}). This introduces a non-negligible probability that the kinetic temperature is lower and the column density is much higher than what we report in Table \ref{table:best_fits} for HD131488. In the case of HD110058, detecting both the optically thick triplet and the optically thin doublet allows us to break this degeneracy. On the other hand, the excitation temperature is unconstrained for both systems as it would need multiple optically thin lines, which are not detected for CI in these systems.

The radial velocity in the heliocentric frame measured for both systems is consistent with the radial velocity of the host star, which is $11.20 \pm 0.81$ km s$^{-1}$ \citep[][]{Iglesias(2018)} for HD110058 and $7\pm2$ km s$^{-1}$ \citep[][]{C.Melis(2013)} for HD131488. This consistency indicates that the observed absorption lines originate from stable gas orbiting the star.

\subsection{CO}

\begin{figure*}
  \includegraphics[width=\linewidth]{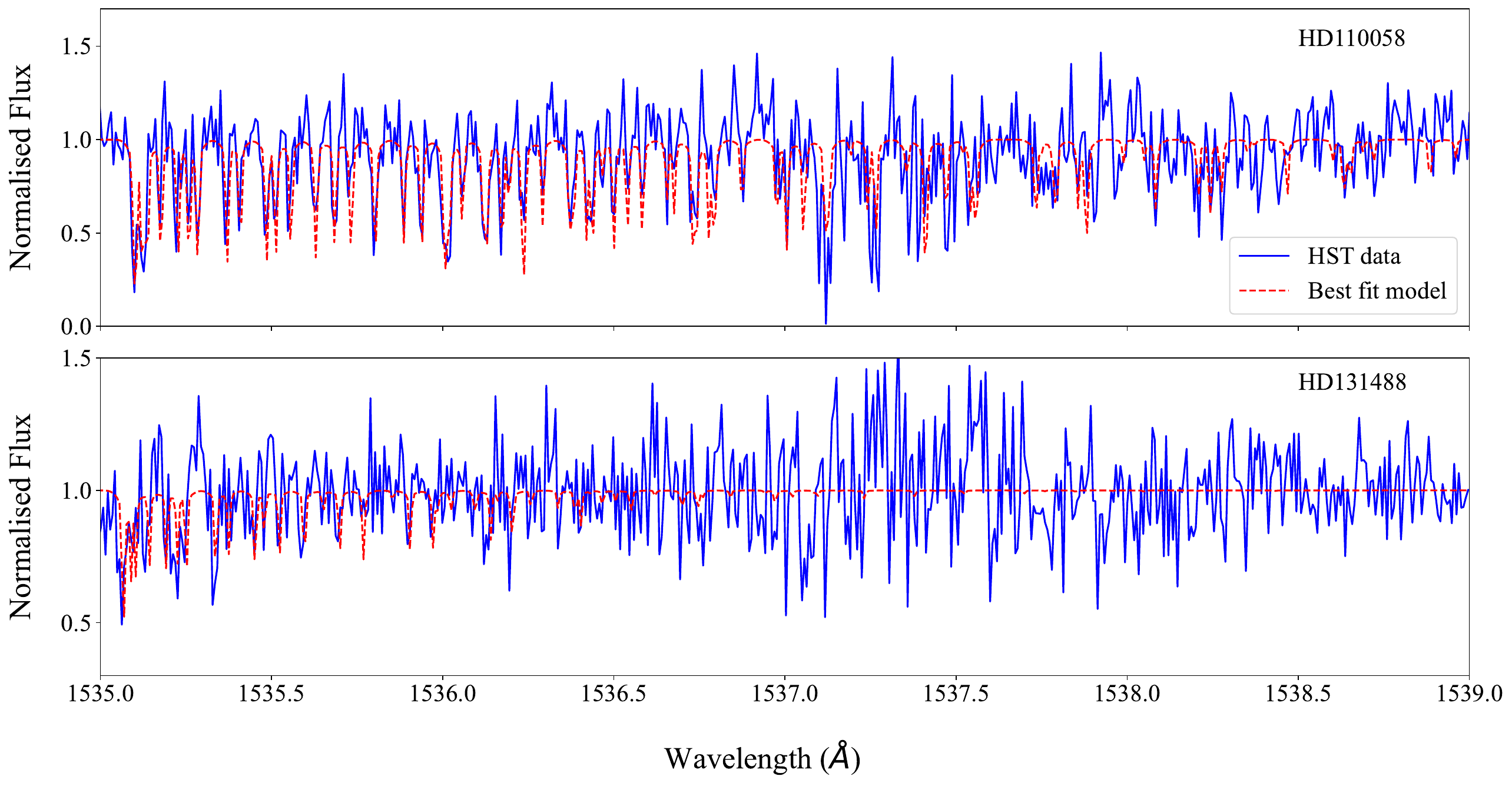}
  \caption{The HST continuum-normalised data  (blue line) for the 1535.09 \AA \ CO band for HD110058 (top) and HD131488 (bottom) were fitted with a sixth-order polynomial to model the continuum, and the resulting best-fit model (red line) was overlaid. All other CO bands for both HD110058 and HD131488 are in the appendix section (see Fig. \ref{fig:HD110_CO} and Fig. \ref{fig:HD131_CO}).}
  \centering
\label{fig:COspectra}
\end{figure*}

\begin{figure}
  \includegraphics[width=\linewidth]{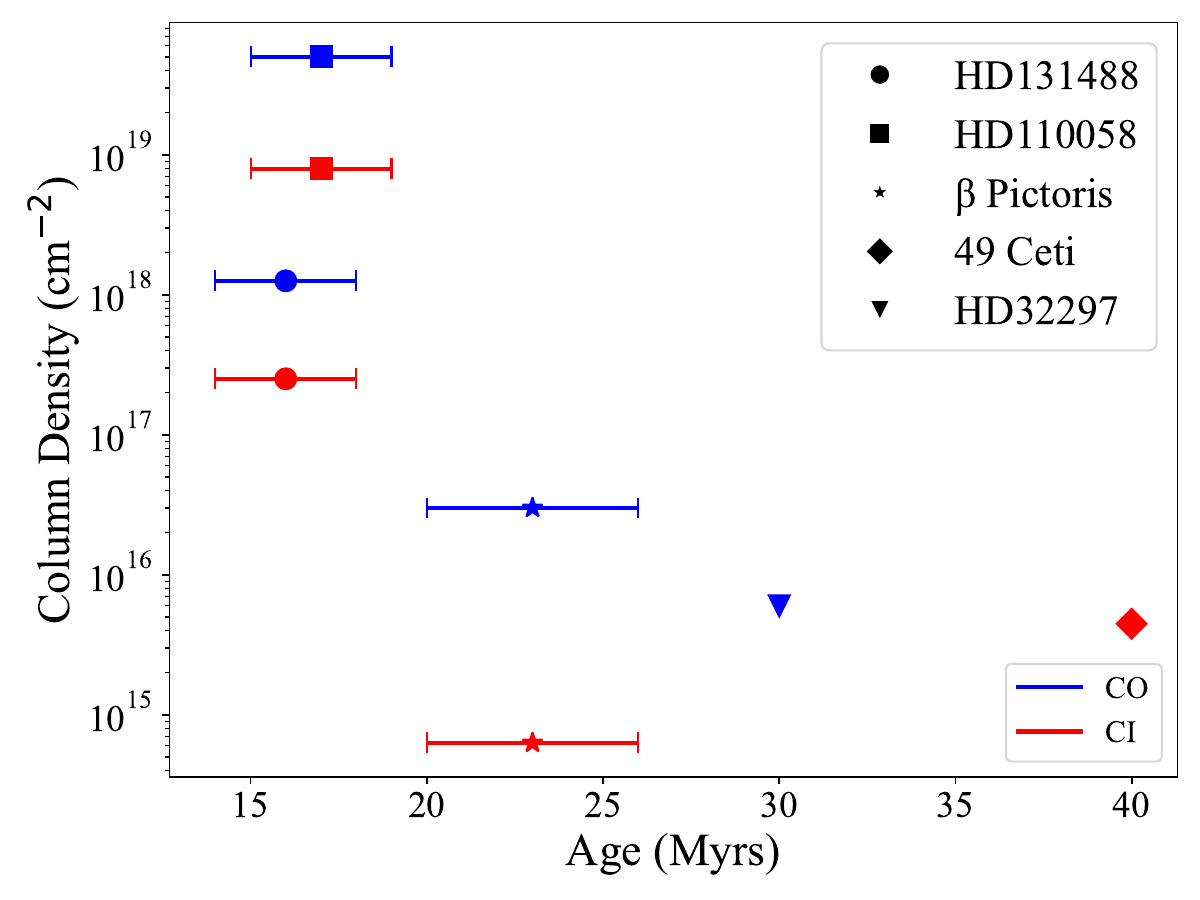}
  \caption{Comparison of all CO (blue) and CI (red) observed in other edge-on exocometary belts. For the 23 Myrs old $\beta$ Pictoris \citep{Mamajek(2014)} both CO and CI column densities are plotted \citep[][]{A.Roberge(2000)}. However, for the 40 Myrs old \citep{Zuckerman(2019)} 49 Ceti only CI was detected \citep[]{A.Roberge(2014)}. For HD32297 the CO column density is an upper limit \citep{K.Worthen(2023)}, where the system age is from \citep{MacGregor(2018)}. The ages for HD110058 and HD131488 are given in \citep{Pecaut(2016)}. The column density uncertainties are too small to be visible on this graph.}
  \label{fig:ComparingDisks}
\end{figure}

Fig. \ref{fig:COspectra} focuses on the fit to the strongly detected CO band at 1535.09 \AA\ for both systems. Plots for the remaining bands can be found in the appendix (Fig. \ref{fig:HD110_CO} and \ref{fig:HD131_CO}). We find CO column densities that are $\sim$ 3 and 4 orders of magnitude higher, respectively, for HD131488 and HD110058 compared to $\beta$ Pictoris (Fig.\ref{fig:ComparingDisks}, \citeauthor{A.Roberge(2000)} \citeyear{A.Roberge(2000)}).

Comparing our CO kinetic temperatures (Table \ref{table:best_fits}) to dust temperatures of 94$\pm$1 K for HD131488 \citep[][]{Vican(2016)} and 112 K for HD110058 \citep[][]{Moor(2017)}, it appears that CO is hotter than the dust temperature for HD110058. Additionally, the dust temperature is likely only representative of micron-sized grains, which dominate the emitting area and are warmer than mm-sized grains. This implies that the discrepancy between CO and the large dust temperature is likely even more significant \citep[e.g.,][]{Matthews(2014)}.

\begin{figure*}
  \includegraphics[width=\linewidth]{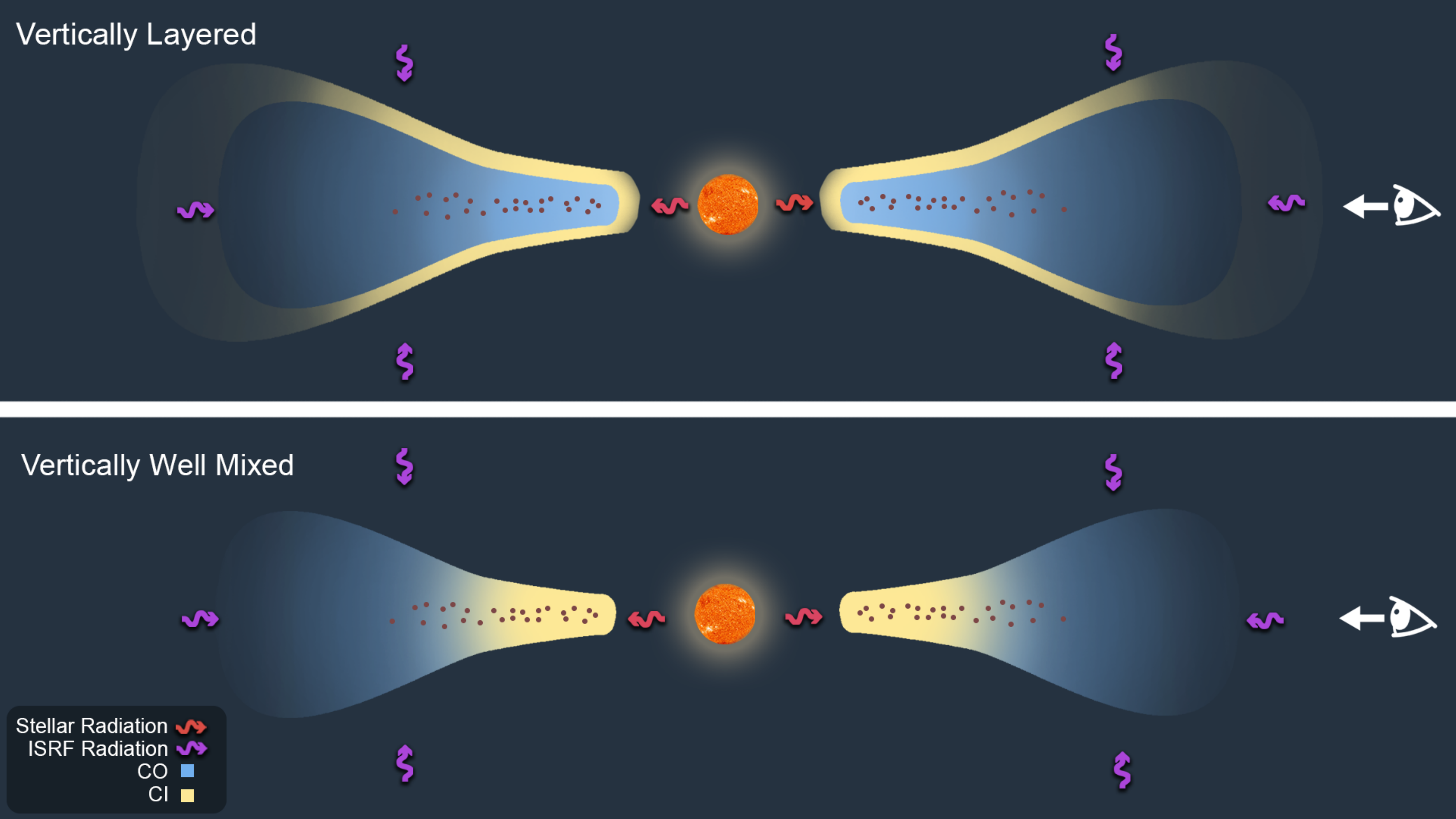}
  \caption{Top panel: In the vertically layered scenario CI is in a vertically thin layer above and below the CO midplane. Bottom panel: In the vertically well-mixed scenario both CO and CI are vertically mixed. Note: This illustration is not intended to accurately reproduce the uncertain radial location of the gas along the line of sight in our targets.}
  \label{fig:Disk layering}
\end{figure*}

We derive consistent kinetic temperatures for CO and CI for HD110058. In the case of HD131488, the degeneracy between column density and kinetic temperature prevents a direct comparison between the CO and CI kinetic temperatures. The consistent CI and CO kinetic temperature for HD110058 may be evidence of co-location, as predicted by gas evolution models if both CI and CO are vertically mixed \citep[][]{S.Marino(2022)} as shown in (Fig. \ref{fig:Disk layering}, bottom panel). However, if CI is in a vertically thin layer above and below the CO midplane as shown in (Fig. \ref{fig:Disk layering}, top panel) \citep[e.g., if vertical diffusion is low;][]{S.Marino(2022)}, our line-of-sight observations may be missing the bulk of the CI and only probe a narrow CI radial layer between the CO inner edge and the star. This assumption is discussed in section 6.

We derive significantly different kinetic and excitation temperatures for HD110058 and potentially (at low significance) for HD131488. This could indicate subthermal populations of the energy levels supporting a second-generation H$_{2}$-poor scenario which is discussed further in section 6.

The accuracy of our CO column density measurements is high for both systems, thanks to the numerous transitions within each of the detected bands. The uncertainty for HD110058 is slightly lower due to the detection of intrinsically weaker absorption lines, stemming from its higher column density. Fig. \ref{fig:COspectra}, \ref{fig:HD110_CO}, and \ref{fig:HD131_CO} demonstrate that our best-fit model (dashed red line) aligns well with the data for the numerous CO bands we analyzed. Small discrepancies in line depths mostly fall within the noise level of each spectrum or could be attributed to uncertainties in the adopted oscillator strengths for each rovibronic transition, or the simplicity of our model assumptions. The oscillator strengths were determined through lab measurements and standard spectroscopic perturbation analysis \citep[][]{Rostas(2000)}. These values generally agree with the observational values, typically within 10\%. Hence, CO transitions exhibiting residuals of approximately 10\% can be explained by the uncertainties on the oscillator strengths.

\renewcommand*{\thefootnote}{\fnsymbol{footnote}}

\begin{table*}
\caption{Best fit values for column density, excitation temperature, kinetic temperature and radial velocity for CI and CO for both systems. The uncertainties are based on the 16th and 84th percentiles of the marginalized distributions. Uncertainties reported for HD131488 CI do not capture the strong degeneracy between the column density and temperature posteriors (see Fig. \ref{fig:HD131_C_corner})}
\label{table:best_fits}
\begin{tabular}{lccccr}
        \hline
		System  & Species & Log Column Density (cm$^{-2}$) & Excitation Temperature (K) & Kinetic Temperature (K) & Radial Velocity (km s$^{-1}$)\\
		\hline
		\rule{0pt}{4ex}    
		HD110058 &  CI & $18.9_{-0.1}^{+0.1}$ &    
            unconstrained & $250_{-40}^{+50}$  & $12.4^{+0.1}_{-0.1}$\\
		\rule{0pt}{4ex}    
            HD131488 & CI & $17.4_{-0.1}^{+0.1}$ & unconstrained  & $1310_{-500}^{+650}$ & $5.20^{+0.1}_{-0.1}$\\
		\rule{0pt}{4ex}   
		HD110058 & CO & $19.7_{-0.1}^{+0.1}$ & $72_{-3}^{+3}$       
            & $193_{-11}^{+12}$  & $12.4^{+0.1}_{-0.1}$\\
		\rule{0pt}{4ex}    
		HD131488 & CO & $18.1_{-0.1}^{+0.2}$ & $45_{-8}^{+8}$     
            & $60_{-10}^{+9}$ & $4.8^{+0.1}_{-0.1}$\\
	    \rule{0pt}{4ex}    
	
\end{tabular}
\end{table*}

\renewcommand*{\thefootnote}{\arabic{footnote}}

\section{Radial Gas Evolution}

We now consider the CO and CI joint column density measurements in the context of the secondary gas model as implemented in the \textsc{exogas \footnote{The \textsc{exogas} code which was used to model radial evolution is available at \url{https://github.com/SebaMarino/exogas}}} python package \citep[][]{S.Marino(2020), S.Marino(2022)}.  In the \textsc{exogas} model, CO gas is continuously released by exocomets within a planetesimal belt, where $\dot{M}_{\mathrm{co}}$ is the rate of CO mass input into the system. This CO gas is photodissociated by the ISRF producing C and O. The gas viscously evolves radially, where the viscosity is parametrized by $\alpha$, eventually forming an accretion disk. These two parameters ($\dot{M}_{\mathrm{co}}$ and $\alpha$) uniquely define the CO and CI mass (and column density) after a certain evolution time \citep[][]{Q.Kral(2019)}. Our joint column density of CO and CI are critical to this model, as the model relies on CI, in particular, to shield CO from otherwise rapid photodissociation, allowing the gas to accumulate producing a shielded second-generation gas disk. This physical mechanism would explain the large CO masses observed in many exocometary disks with ALMA, including HD110058 \citep[][]{A.Hales(2022)} and HD131488 \citep[][]{Smirnov-Pinchukov(2022)}. In section 4.1, we outlined our assumptions for the \textsc{exogas} model, while in section 4.2 we describe the addition of stellar photodissociation. To summarize our findings, we determined that this addition remains inconsequential for typical CO-rich disks and host stars, even at distances as close as a few au, due to the substantial radial optical depths. The potential ramifications of introducing the stellar photodissociation rate are further explored in section 4.3.

\subsection{Model assumptions}

In line with \citet[][]{S.Marino(2020)}, we make some assumptions to simplify the modelling. Firstly, we set the C ionisation fraction to 0.1, which is reasonable for shielded disks when the carbon densities are expected to be high \citep[][]{S.Marino(2020)}. Secondly, the effect of radiation pressure, as a mechanism to remove CI from the system was ignored. This is justified as the continuous production of CI in the secondary gas scenario causes CI gas to reach densities capable of self-shielding from stellar radiation preventing blowout \citep[][]{Q.Kral(2017)}. The original model did not include the stellar contribution to CO photodissociation, as it was deemed negligible compared to the ISRF for stellar luminosities below 20 L$_{\odot}$ and debris disc sizes of $\sim$100 au \citep[][]{S.Marino(2020)}. However, as the belt of one of our targets (HD110058) is located much closer to the star we incorporated the effect of stellar photodissociation into \textsc{exogas} as described in section 4.2.

Our HST observations probe the gas column along the line of sight to the star. For simplicity, we assume the systems to be perfectly edge-on, consistent with the current ALMA CO results for HD110058 \citep[][]{A.Hales(2022)} and HD131488 \citep[][]{Moor(2017)}.  In our \textsc{exogas} modelling, we also assumed that vertical diffusion is strong and that the gas is well mixed vertically (Fig. \ref{fig:Disk layering}, bottom panel), with both CO and CI present in the midplane and therefore traced by our observations. If instead, the gas were vertically layered (Fig. \ref{fig:Disk layering}, top panel), with CI in a thin layer above and below the CO-rich midplane, then our measured CI/CO ratio will be a lower limit to the overall CI/CO abundance.  Finally, to translate surface densities $\sum$ (in the vertical direction) produced by the model into radial column densities, we assume that the disk has a constant aspect ratio of 0.1  (strong vertical diffusion), where the CI and CO gas is vertically mixed throughout the midplane.

\begin{table*}
\caption{Best fit values determined from SED fits.}
\label{table:stellar_fitting}
\begin{tabular}{lcccccr}
        \hline
		Star & Temperature (K) & Log g & Fe/H & Distance (pc) & Radius   (R$_{\odot})$ & A$_{\mathrm V}$ (magnitudes) \\
		\hline
		\rule{0pt}{4ex}    
		HD110058 & $7990_{-50}^{+120}$& $3.54_{-0.01}^{+0.41}$&$-0.3_{-0.2}^{+0.2}$&$130.2_{-0.2}^{+2.0}$&$1.55_{-0.04}^{+0.02}$& \multicolumn{1}{c}{$0.03_{-0.01}^{+0.06}$}\\
		\rule{0pt}{4ex}      
		HD131488 & $8700_{-210}^{+160}$& $3.80_{-0.16}^{+0.38}$&$-0.1_{-0.12}^{+0.17}$&$152.4_{-0.2}^{+3.2}$&$1.60_{-0.02}^{+0.03}$&
            \multicolumn{1}{c}{$0.01_{-0.02}^{+0.04}$}\\

\end{tabular}
\end{table*}

\subsection{Stellar photodissociation}

To include the stellar photodissociation rate, we need to determine the UV flux from the star at wavelengths where CO photodissociation occurs $\sim$900-1100 \AA. These wavelengths are not covered by our COS data, which detects the star down to $\sim$1220 \AA\ before the continuum is not detected. Therefore, we need to use a stellar model to extrapolate the flux to the desired wavelengths. To do so we fit optical and near-IR photometry (Table \ref{table:photometry_data}) combined with precise Gaia parallaxes for both stars using the astroARIADNE\footnote{Available at \url{https://github.com/jvines/astroARIADNE}} package \citep[][]{Vines(2022)}. In summary, astroARIADNE uses Bayesian model averaging to fit up to 6 different stellar atmospheric model grids, to obtain effective temperature, surface gravity, metallicity, distance, radius and V-band extinction.

The best-fit models for HD110058 and HD131488 are Castelli and Kurucz models, with 50$\pm$34 percentiles of the posterior probability distribution for each parameter shown in Table 4. Fig. \ref{fig:stellar_HD110} and \ref{fig:stellar_HD131} (top panels), show the best-fit model (orange line) overplotted with the fitted optical and near-IR photometry (coloured data points). To account for potential inaccuracies in the model extrapolation from the fitted optical data to $\sim$900-1100 \AA, we rescale the continuum of the best-fit models to match the observed stellar continuum at the shortest wavelengths $\sim$1220 \AA \ of our COS data (Fig. \ref{fig:stellar_HD110} and \ref{fig:stellar_HD131}, bottom panels). The stellar models underestimate the HST spectrum in this far-UV region by factors of 9 and 20 for HD110058 and HD131488 at a wavelength of $\sim$1250 \AA. The same effect was observed for a similarly young A star $\beta$ Pictoris \citep{L.Matra(2018b)}, where this is likely due to stellar activity \citep{Deleuil(2001), Bouret(2002)}. In Figs. \ref{fig:stellar_HD110} and \ref{fig:stellar_HD131}, the COS HST data and model were scaled to 22 au (HD110058) and 35 au (HD131488), respectively. These distances correspond to the peak dust surface density for HD110058 \citep[][]{A.Hales(2022)} and the inner gas disk radius for HD131488 \citep[][]{Smirnov-Pinchukov(2022)}. These distances are just representative distances, whereas the code includes the distance-dependent stellar flux. 

To include the stellar photodissociation rate in the code, we first calculate the unshielded stellar photodissociation rate by combining our stellar model fluxes with CO photodissociation cross-sections from the Leiden\footnote{Cross-sections taken from \url{https://home.strw.leidenuniv.nl/~ewine/photo/display_co_42983b05e2f2cc22822e30beb7bdd668.html}} database.  Due to the radial distribution of gas in debris disks, we expect gas closer to the star to provide shielding to gas that is further from the star. At every radial location, we determine the radial column density of CO and CI between the star and the given radius, which has contributions from all the disk locations at shorter radii. To account for CO self-shielding and CI shielding, at each radius we use the column density of CI and CO interior to calculate the shielding function, where the CO self-shielding functions are given by

\begin{equation}
f_{\text{self-shielding}} = \frac{\int \phi_{\lambda} \sigma_{\text{CO}}(\lambda)\exp[-N_{\text{CO}}\sigma_{\text{CO}}(\lambda)] \,d\lambda}{\int \phi(\lambda) \sigma_{\text{CO}}(\lambda) d\lambda
}
\end{equation}

where, $\phi_{\lambda}$ (photons s$^{-1}$ nm$^{-2}$) is the stellar flux at the radius in question, $\sigma_{\text{CO}}(\lambda)$ (cm$^{2}$) are the CO cross sections, $N_{\text{CO}}$ (cm$^{-2}$) is the CO column density in the radial direction interior to the radius in question, finally $\int \phi(\lambda) \sigma_{\text{CO}}(\lambda) d\lambda$ is the unshielded stellar photodissociation rate (s$^{-1}$). Similarly, the CI shielding functions are given by

\begin{equation}
f_{\text{carbon-shielding}} = \frac{\int \phi_{\lambda} \sigma_{\text{CO}}(\lambda) \exp[-N_{\text{CI}}\sigma_{\text{CI}}(\lambda)] \,d\lambda}{\int \phi(\lambda) \sigma_{\text{CO}}(\lambda) d\lambda
}
\end{equation}

where $N_{\text{CI}}$ (cm$^{-2}$) is the CI column density in the radial direction interior to the radius in question. Finally, we combine the relevant shielding functions with the unshielded stellar photodissociation rate to get the shielded CO photodissociation rate. In conclusion, at every time step in the \textsc{exogas} model evolution, the photodissociation rates due to stellar radiation are computed at every radius, including the shielding effects described above.

\subsection{Impact of stellar photodissociation}

\begin{figure*}
  \includegraphics[width=\linewidth]{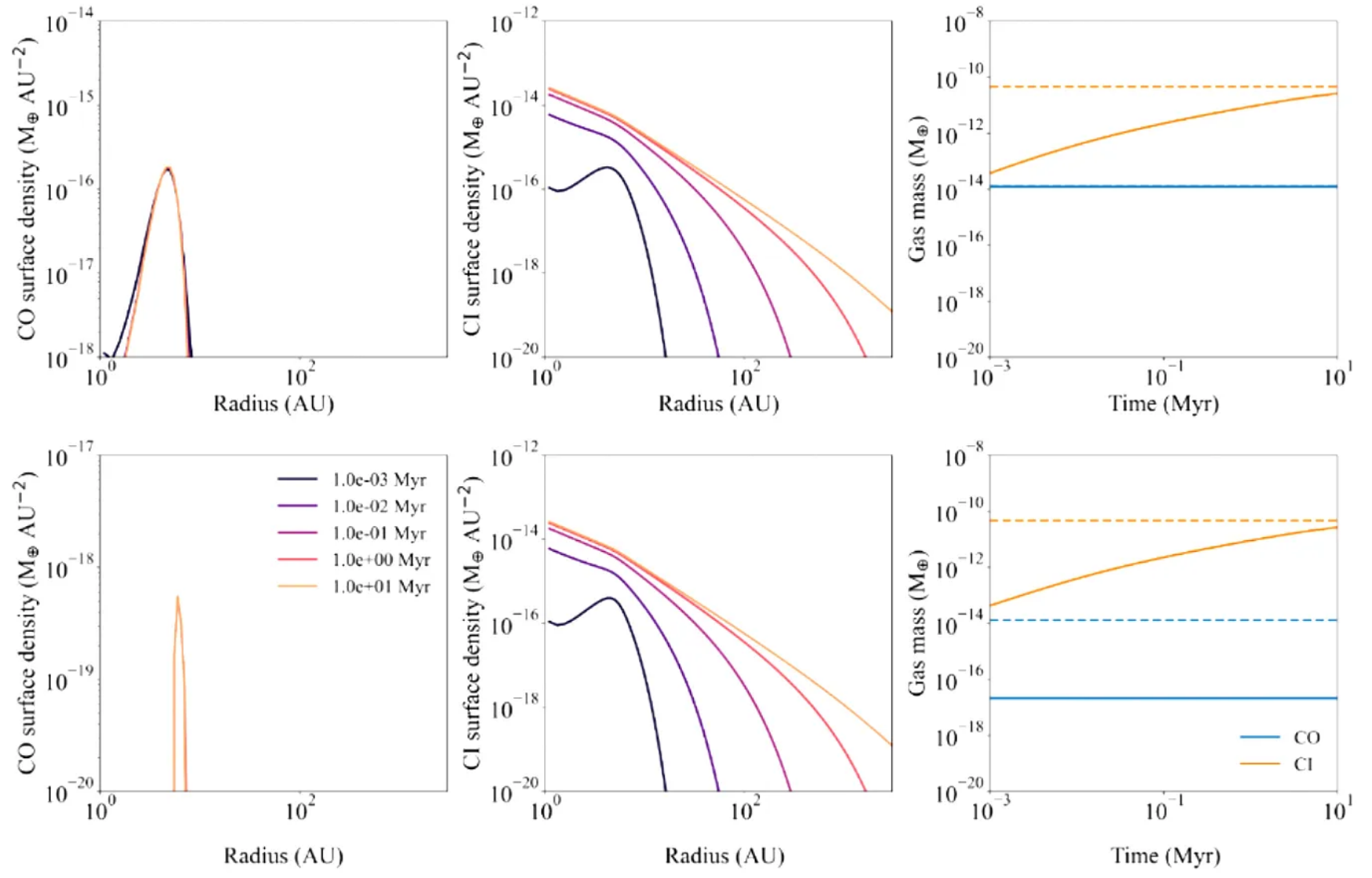}
  \caption{Simulations for the low CO input rate and high viscosity scenario, the top panels show the scenario where only the effect of the ISRF is included in the model. The bottom panels included the effect of the ISRF and the star. The left and middle panels show the surface densities of CO and CI, respectively, as a function of time. The right panels show the total gas mass for both CO (blue continuous line) and CI (orange continuous line) as a function of time. Additionally, the blue dashed line represents the CO mass expected in the unshielded steady state, i.e. where CO production is balanced by destruction through unshielded photodissociation in the vertical direction. The orange dashed line represents the CI mass expected in steady state in a scenario where CI production is through unshielded photodissociation, and CI destruction is through accretion onto the star.}
  \label{fig:unshielded}
\end{figure*}

\begin{figure*}
  \includegraphics[width=\linewidth]{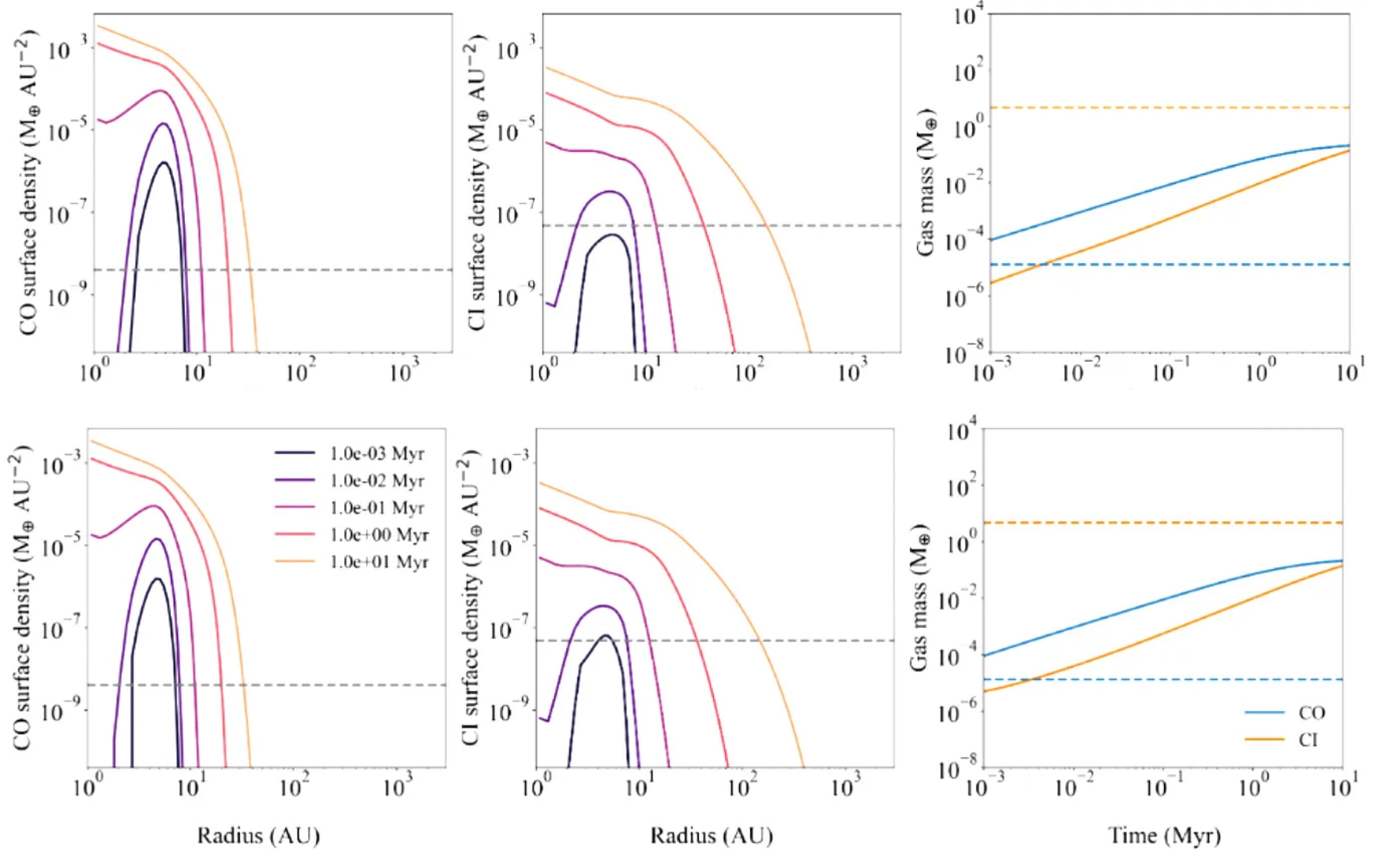}
  \caption{Simulations for the low CO input rate and high viscosity scenario, the top panels show the scenario where only the effect of the ISRF is included in the model. The bottom panels included the effect of the ISRF and the star. The left and middle panels show the surface densities of CO and CI, respectively, as a function of time. The right panels show the total gas mass for both CO (blue continuous line) and CI (orange continuous line) as a function of time. Additionally, the blue dashed line represents the CO mass expected in the unshielded steady state, i.e. where CO production is balanced by destruction through unshielded photodissociation in the vertical direction. The orange dashed line represents the CI mass expected in steady state in a scenario where CI production is through unshielded photodissociation, and CI destruction is through accretion onto the star. }
  \label{fig:shielded}
\end{figure*}

To illustrate the impact of the star, we consider two gas evolution scenarios\footnote{These two extreme cases achieve very low and very high shielding.} of a low CO input rate, high viscosity scenario ($\dot{M}_{\mathrm{co}}$ = $10^{-10}$ M$_\oplus$ Myrs$^{-1}$, $\alpha$ = $10^{-1}$), and a high CO input rate, low viscosity scenario ($\dot{M}_{\mathrm{co}}$ = $10^{-1}$ M$_\oplus$ Myrs$^{-1}$, $\alpha$ = $10^{-3}$). We consider a system with an exocometary belt at 5 au from a central A-type star\footnote{We also assume a stellar mass of 1.8 M$_{\odot}$, a luminosity of 9 L$_{\odot}$, a debris disk radius of 5 au, and an FWHM of 2.5 au.}, where the stellar UV flux will be much larger than that from the ISRF. We trace the radial evolution of the gas using \textsc{exogas} for [0.001,0.01,0.1,1, 10] Myrs (colours in Fig. \ref{fig:unshielded} and Fig. \ref{fig:shielded}).

We start by considering the simple behaviour in the low CO input and high viscosity scenario when only the effect of the ISRF is included in the model (Fig. \ref{fig:unshielded}, top panels). CO is released in the belt and quickly photodissociated, producing CI and O. The CO gas quickly reaches a steady state (top right panel), at which point the CO mass is $\dot{M}_{\mathrm{co}}$ $\times$ 130 yrs (the unshielded CO lifetime). Over a few viscous timescales ($10$ Myrs for $\alpha$ = $10^{-1}$ in this model), CI viscously spreads forming an accretion disk, eventually reaching a steady state (top right panel), where the CI destruction rate by accretion onto the star is equal to the CI production rate by CO photodissociation. 

Next, we examine the same scenario, but this time we include the effect of stellar photodissociation in addition to the ISRF  (Fig. \ref{fig:unshielded}, bottom panels). The steady-state CO mass reached (bottom right panel) is lower compared to the ISRF-only case. For the same CO production rate, the photodissociation rate is significantly higher due to the stellar contribution, hence the resulting steady-state CO mass is lower (solid vs dashed blue lines). After CO is in steady state the CI gas follows the same evolution as before since the destruction rate of CO (equal to its production rate) is the same. Including the star has an important effect on the radial distribution and mass evolution of CO in the very low CO input rate regime, where the disk remains unshielded radially from both the star and the ISRF, and vertically from the ISRF. 

We now consider the high CO input rate, and low viscosity scenario, for the ISRF-only and the (combined ISRF and stellar) cases (Fig. \ref{fig:shielded}, top and bottom respectively). In this high CO input rate regime, the CO rapidly reaches a level where it is shielded from the star in the radial direction (along the midplane). This happens much faster than shielding from the ISRF in the vertical direction, due to the higher column densities along the disk midplane compared to perpendicular to it. As a result, stellar photodissociation rapidly becomes negligible compared to ISRF photodissociation in the vertical direction. This implies that, as shown in Fig. \ref{fig:shielded} (top vs bottom) the gas evolution in the two models is almost undistinguishable in the shielded, high CO input rate scenario. In conclusion, we have modified the \textsc{exogas} evolution model to include stellar photodissociation and found that this is only important at a few au around A-type stars, for very low CO input rates which are unlikely to lead to detected CO levels.

\section{Application to HD110058 and HD131488}

We now apply the updated model to simulate the evolution of gas for HD110058 and HD131488. For HD110058, we assume a stellar mass of 1.8 M$_{\odot}$, a stellar luminosity of 9 L$_{\odot}$, a debris disk where the surface density peaks at 22 au, with an inner and outer FWHM of 6 au and 82 au \citep[][]{A.Hales(2022)}. For HD131488, we assume, a stellar mass of 1.88 M$_{\odot}$, a stellar luminosity of 13.1 L$_{\odot}$, a debris disk radius of 88 au, FWHM of 44 au  \cite{Moor(2017)}. Additionally, we assumed a C ionisation fraction of 0.1 for both systems. We ran a grid of 31$\times$21 models for different $\dot{M}_{\mathrm{co}}$ and $\alpha$ values \footnote{We set reasonable values for the CO input rate and $\alpha$ based on prior studies. The $\alpha$ parameter's upper limit is guided by high inferred values (i.e., $\alpha$ > 0.1) in the $\beta$ Pic gas disc \citep{Q.Kral(2019)} and the lower limit is set by systems with a smaller carbon ionization fraction ($\alpha$ > 10$^{-3}$) \citep{Kral&Latter(2016)}. Reasonable CO input rates are determined from modelling results of 12 debris disks \citep{Q.Kral(2016)}.} calculating the total CO and CI column density after 10 Myrs\footnote{10 Myrs is frequently used as the transition point between protoplanetary disks and debris disks \citep{Wyatt(2015)}.} of evolution for both systems. The outputted CO and CI column densities are shown in the colour scale in Fig. \ref{fig:contour_plot}. To help the reader to understand the variation in column density with $\dot{M}_{\mathrm{co}}$ and $\alpha$ shown in Fig. \ref{fig:contour_plot}, we plot the behaviour of CI column density as a function of the viscous parameter $\alpha$ in Fig. \ref{fig:Mass_N}. Therefore, Fig. \ref{fig:Mass_N} can be interpreted as a horizontal cross-section of Fig. \ref{fig:contour_plot}.  

\subsection{Low CO input rate ($\dot{M}_{\mathrm{co}}$)}

We start by considering the low CO input rate case, for high $\alpha$ parameter, which corresponds to the darkest region in the colour scale (bottom right in each of Fig. \ref{fig:contour_plot} panels). In this scenario, the CO quickly reaches a steady state where the CO mass is equal to $\dot{M}_{\mathrm{co}}$ $\times$ 130 yrs. The CO is unshielded from the ISRF in the vertical direction (but shielded from the star in the radial direction, see section 4.3). The high $\alpha$ implies a short viscous timescale that allows CI to rapidly spread radially, evolving into an accretion disk profile. This leads to a steady state where the rate of CI production by CO photodissociation equals the rate of CI destruction by accretion.

We then move to examine the gas evolution as the $\alpha$ value decreases (right to left in all panels in Fig. \ref{fig:contour_plot}). Initially, the gas evolution is the same as the vertical unshielded case. However, as the CO gas photodissociates, the newly produced CI gas accumulates at a narrow range of radii due to the low viscosity. Eventually, enough CI gas has accumulated for CI shielding to occur,  after which the CO photodissociation timescale increases and the CO gas mass rises to a new shielded steady state. Therefore, as we move from high to low alpha values, we see an increase in the expected CO column density in  Fig. \ref{fig:contour_plot}. The CI mass dependence on $\alpha$ at low CO input rates (Fig. \ref{fig:Mass_N}, solid lines) is driven largely by the viscous accretion timescale. At high  $\alpha$ values, viscous timescales are short causing CI to rapidly start accreting onto the star. Decreasing  $\alpha$ lengthens the viscous timescale, and therefore the time required for CI to start accreting. At low enough $\alpha$ values, we observed a plateau in CI gas mass because the viscous timescale becomes significantly longer, implying that most of the CI produced in 10 Myr is yet to be destroyed by accretion. 

Interestingly, this increase and then plateau in the CI mass turns into a peak when expressed as radial column density as shown in Fig. \ref{fig:Mass_N}. This is because as $\alpha$ increases the CO spreads inwards more and photodissociates, releasing CI gas at smaller radii. Adding CI at shorter radii increases the radial column density more than if CI was added at larger radii, for a disk with an assumed aspect ratio $h=H/r$ that is constant with radius. If we divide our model disk into rings of constant width $\Delta$r, the radial column density in that ring is related to its vertical surface density $\Sigma$ through 
\begin{equation}
N = \Sigma \frac{\Delta r}{2H}
\end{equation}
where $H$ is the scale height of the disk. As in our model, we assume the aspect ratio to be constant with the radius, assuming $H=hr$, which allows us to obtain,
\begin{equation}
N\propto \frac{M}{r^{2}}
\end{equation}
Therefore, the radial column density is most sensitive to the addition of gas in the inner regions, which is favoured for higher $\alpha$ values (more radial spreading). This produces the observed column density peak compared to the mass plateau as a function of $\alpha$. 

In this low $\dot{M}_{\mathrm{co}}$  regime we observe the same behaviour for both of our systems Fig. \ref{fig:Mass_N} (blue vs red), and Fig. \ref{fig:contour_plot} (top vs bottom). However, we note that the peak column density for CI in HD131488 is shifted to higher $\alpha$ values compared to HD110058. This is because for a fixed $\alpha$, the viscous timescale increases with radius. As the gas in HD131488 is inputted at significantly larger radii, it takes longer for CI to form an accretion disk and reach a steady state. Therefore, to achieve an accretion disk for CI for a fixed evolution time of 10 Myrs, the $\alpha$ required at larger radii is higher.
 
\subsection{High CO input rate ($\dot{M}_{\mathrm{co}}$)}

Next, we examine the evolution of gas for high CO input rates and high $\alpha$ values (top right for all panels in Fig. \ref{fig:contour_plot}). As the gas spreads the threshold for CO self-shielding is reached first, and then CI shielding begins shortly after at which point the CI shielding becomes more dominant. After 10 Myrs our systems have formed an accretion disk but have not fully reached a steady state in this high CO input rate, high $\alpha$ region of parameter space. 

As we decrease the $\alpha$ parameter (moving from right to left in Fig. \ref{fig:contour_plot}), CI gas spreads out less and starts to accumulate, as the CI gas is not being significantly accreted onto the star at low $\alpha$ rates compared to the high $\alpha$ case. This causes the CI mass to increase for low $\alpha$ values and then plateaus e.g. as shown in Fig. \ref{fig:Mass_N} (blue and red dashed lines for HD110058 and HD131488, respectively).
Similarly to the low CO input rate, the plateau in CI gas mass for the high CO input rate corresponds to a peak in CI column density in Fig. \ref{fig:Mass_N}.

\subsection{Comparison with HST data}

\begin{figure*}
  \includegraphics[width=\linewidth]{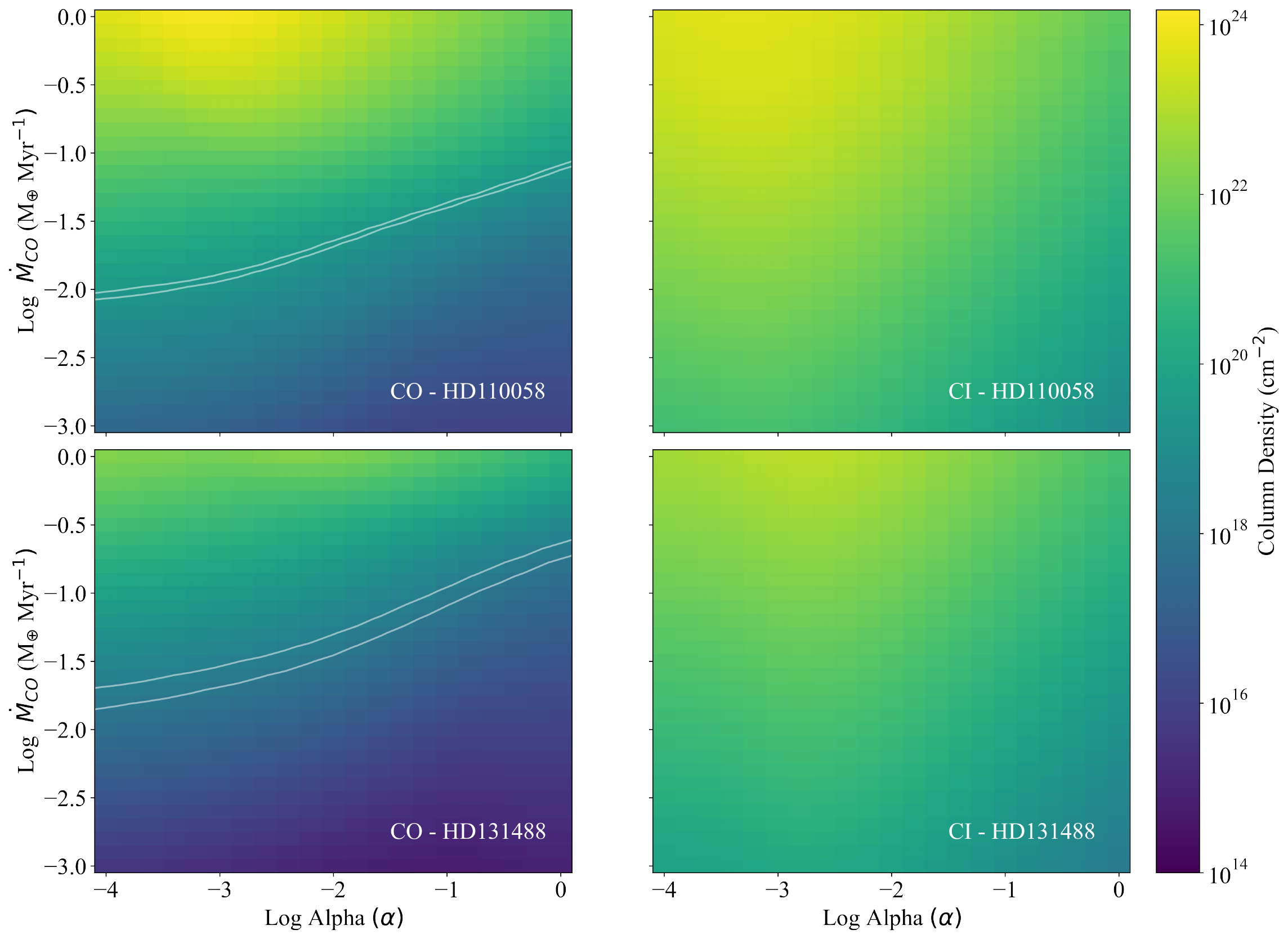}
  \caption{Colour plots of outputted CO (left plots) and CI (right plots) column densities for HD110058 and HD131488 after evolving for 10 Myrs. The best-fit values for CO are shown on the left panels (white contours), but the best-fit values for CI are not shown because they are outside the range of the graph.}
  \label{fig:contour_plot}
\end{figure*}

\begin{figure}
  \includegraphics[width=\linewidth]{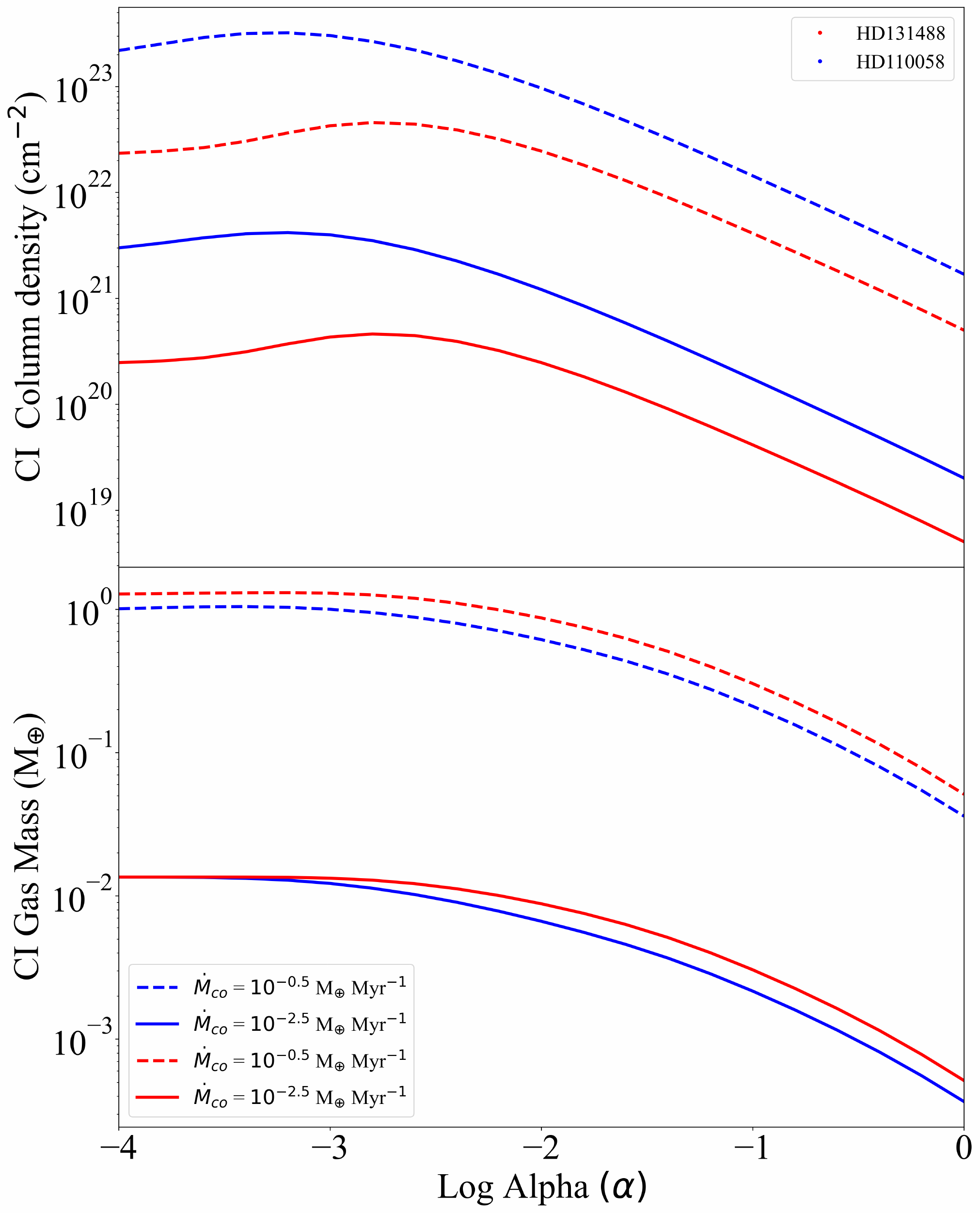}
  \caption{Column densities (top) and gas masses (bottom) for CI after the HD110058 system (blue lines) and the HD131488 (red lines) have been evolved for 10 Myrs as a function of $\alpha$ for 2 different $\dot{M}_{\mathrm{co}}$ of $10^{-2.5}$ M$_\oplus$ Myrs$^{-1}$ (solid), $10^{-0.5}$ M$_\oplus$ Myrs$^{-1}$ (dashed). This graph plots the evolution of CI gas for different CO input rates and can be thought of as horizontal sections of the CI colour plot Fig.\ref{fig:contour_plot}.}
  \label{fig:Mass_N}
  
\end{figure}

For comparison, in Fig. \ref{fig:contour_plot} we over-plot the distribution (16th-84th percentile) of CO column densities (white contours) that best fit the HST data (Table \ref{table:best_fits}). The best-fit values for CO require low alpha parameters for low CO input rates, and high alpha parameters are required for high CO input rates as expected. Interestingly, the best-fit values for CI in both systems are outside the lower end of the colour scale values for both systems, which is why we have not shown them in Fig. \ref{fig:contour_plot}. The fact that the CO and CI constraints (contours) do not cross anywhere in the ($\dot{M}_{\mathrm{co}}$, $\alpha$) parameter space considered in this study, indicates that there is no combination of CO input rates and viscosity that can simultaneously explain our CO and CI column densities, measured with HST. We find that the model cannot reproduce, and systematically over-predicts the CI/CO ratio for the reasonable values of $\dot{M}_{\mathrm{co}}$ and $\alpha$ we explored. For HD131488, we note that even for much lower kinetic temperatures and higher column density values than our nominal 1-$\sigma$ uncertainties, which are not excluded at high significance by the data (Fig.\ref{fig:HD131_C_corner}), the observed CI column density would still be much lower than predicted by the model.

\section{Discussion}

Our modelling results show that the column density of CI required by the secondary gas model to explain the column density of observed CO is much higher than measured in our HST observations. In other words, the model would imply a CI/CO ratio that is too high compared to our observations for both systems. Hence, the vertically well-mixed, second-generation gas release scenario is missing some important physics or chemistry, one of our assumptions is wrong, or the gas is not of second generation around these systems. This result is in agreement with recent results from ALMA CO and CI mass measurements \citep[][]{Cataldi(2023), A.Hales(2022)}, which also indicates that this second-generation model overpredicts CI, in consensus with our findings. There are a few possible ways that the second-generation origin scenario could reconcile with available data as suggested in \citet[][]{Cataldi(2023)}.

The simplest way to reduce the CI/CO ratio would be to add a removal process for CI other than accretion, the only process considered by our modelling in this paper. Another possibility is chemistry reforming CO from CI through a chain of reactions involving  H$_{2}$-bearing molecules in the gas phase \citep[][]{Higuchi(2017)}. The necessary H$_{2}$ could be provided, for example, by exocometary water in a secondary scenario. Removal of CI by radiation pressure could also in principle alleviate the issue but is unlikely as even low levels of CI can prevent blowout by self-shielding \citep[][]{Q.Kral(2017)}. Another shielding agent beyond CI and CO itself may be responsible for shielding the high observed CO levels while maintaining a low CI/CO ratio; however, currently, there are no obvious candidate species as H$_{2}$ isn't expected to exist in second-generation (exocometary) gas in debris disks at levels which would allow shielding to occur. It is also important to consider that secondary gas may not be produced by a steady-state collisional cascade. Recent studies propose that the gas could be produced by a giant collision or a tidal disruption, which could help explain the discrepancies between the model predictions and observations \citep{Cataldi(2020)}.

Another explanation for the discrepancy between our simulations and the observed CI column density could be that CO and CI gas are not well mixed vertically throughout the midplane (Fig. \ref{fig:Disk layering}, bottom panel), but are instead vertically layered (Fig. \ref{fig:Disk layering}, top panel) with CI in a thin layer above and below the CO midplane \citep[][]{S.Marino(2022)}. If this is the case, a lower midplane CI/CO ratio than in the well-mixed case is predicted, potentially reproducing the observed low CI/CO ratios. This is because if CI is all in a narrow vertical layer the shielding would be enhanced, resulting in CO photodissociating on a longer timescale.This suggests that assuming vertical layering could resolve the discrepancy between the simulations and our best-fit values. This potential vertically layered solution was originally supported by modelling of the ALMA CO emission data \citet[][]{A.Hales(2022)}, but recently disfavoured by CI ALMA emission observations \citet[][]{Cataldi(2023)} showing low CI/CO ratios, as found here, when probing the entirety of the gas disk in emission with ALMA, rather than just the line of sight as probed by HST.

It is worth noting that if the gas were vertically layered and CI was largely in a thin layer above and below the midplane (Fig. 3, top panel), our observations (assuming they probe the midplane) would miss most CI gas. In that case, HST absorption spectroscopy would only be probing the thin CI layer between the star and the inner edge of the CO disk and between the outer edge of the CO disk and the interstellar medium. Therefore, it would be crucial to vertically resolve CO and CI emission with ALMA to definitively rule out this vertically layered scenario. However, for HD110058 the observed CI and CO along the line of sight to the star are consistent with having the same kinetic temperature. This would instead indicate that CI and CO may be co-located, and argue against a vertically (and radially) layered scenario.

Additionally, the assumption of a perfectly edge-on disk also has fundamental implications for our results. If the disk is not viewed edge-on (e.g. inclination $\sim$ 80$^{\circ}$ - 85$^{\circ}$) such that the line of sight goes at 1-2H above the midplane, the CI/CO ratio is unchanged with height if CO and CI are vertically well mixed \citep{S.Marino(2022)}. However, if CO and CI are vertically layered, CI/CO increases above the midplane, therefore our measured midplane CI/CO ratio would be an upper limit. 

In our model, we adopted a C ionization fraction of 0.1 however we also tested for a higher ionisation fraction of 0.5. We found that assuming a higher ionisation fraction resulted in a lower model CI/CO ratio which is slightly closer to the low CI/CO ratio observed. However, assuming a higher ionisation fraction does not produce a CI/CO ratio low enough to change our conclusions.

Finally, we cannot rule out a primordial origin for the gas in our two systems, with H$_{2}$ acting as the main CO shielding agent, as the level of H$_{2}$ present in these systems remains unknown and difficult to observe. However, the higher observed kinetic temperatures compared to excitation temperatures for HD110058 suggest an NLTE scenario and therefore low collider densities, as would be expected in an H2-poor scenario (secondary gas scenario) \citep{L.Matra(2015)}.

In conclusion, our HST findings of a low CI to CO column density ratio in the CO-rich debris disks of HD110058 and HD131488 are fully independent of but in line with recent ALMA findings of low CI to CO mass ratios in similar disks, which cannot be explained by our current simple model of second-generation gas evolution after production from a collisional cascade. This could indicate a different origin scenario, but could also be explained in a second-generation context by additional CI removal mechanisms, shielding agents and/or low vertical diffusion producing a vertically thin CI layer above and below the CO midplane.

\section{Summary}

In this paper, we presented HST absorption spectroscopy observations of two young edge-on debris disks (HD110058 and HD131488), covering several CI electronic transitions and CO rovibronic bands with a range of oscillator strengths, allowing us to constrain the CI and CO column densities and average line-of-sight kinetic temperatures. To interpret the jointly measured CO and CI column densities, we used the \textsc{exogas} second-generation gas model \citep[][]{S.Marino(2020)} to simulate the release and radial evolution of gas produced within debris disks around the two host stars. Our main findings are as follows:
           
\begin{itemize}
\item We detect very high levels of CI and CO of $18.9_{-0.1}^{+0.1}$ cm$^{-2}$ and $19.6_{-0.1}^{+0.1}$ cm$^{-2}$ for HD110058 and $17.4_{-0.1}^{+0.1}$ cm$^{-2}$ and $18.0_{-0.1}^{+0.2}$ cm$^{-2}$ for HD131488. The observed CO  column densities are $\sim$ 3-4 orders of magnitude larger than observed in the $\beta$ Pictoris disk.
\item We find CO kinetic temperatures higher than the dust temperatures inferred from the SED for HD110058. Additionally, we find consistent kinetic temperatures for CO and CI for HD110058. For HD131488, the degeneracy between column density and kinetic temperature prevents a direct comparison between the CO and CI kinetic temperatures. The consistent CI and CO kinetic temperature for HD110058 may be evidence of the co-location of CI and CO.
\item We find that the impact of stellar radiation on photodissociation radially is minimal compared to the influence of ISRF photodissociation in the vertical direction for high exocometary release rates as expected in detected systems. This is due to the rapid development of an optically thick radial column of CO and CI, which prevents the penetration of UV radiation in the radial direction.
\item We find that the secondary gas model implies a CI/CO ratio that is too high compared to our HST observations for both systems. Therefore, either the gas is not mixed throughout the midplane as assumed, one of our assumptions is wrong or the current secondary gas model is not accurate. 

\end{itemize}

In conclusion, more complex models are likely needed to explain the observed low CI/CO ratios independently determined by HST UV line-of-sight column density here. Future observations should thoroughly test these models, for example by determining the vertical structure of CI and CO, which significantly influences shielding effectiveness. Furthermore, these observations may yield evidence for other gas species, such as H$_{2}$, that could potentially serve as shielding agents, either through direct measurement or indirect inference. This will be crucial to confirm or rule out gas origin and/or evolution scenarios and provide us with a better understanding of the role of gas in young debris disks in the latest stages of planet formation.

\section{Acknowledgements}
AB and LM acknowledge research support by the Irish Research Council under grants
GOIPG/2022/1895 and IRCLA/2022/3788. SM is supported by a Royal Society University Research Fellowship (URF-R1-221669). AMH is supported by a Cottrell Scholar Award from the Research Corporation Science Advancement. This research is based on observations made with the NASA/ESA Hubble Space Telescope obtained from the Mikulski Archive for Space Telescopes (MAST) at the Space Telescope Science Institute, which is operated by the Association of Universities for Research in Astronomy, Inc., under NASA contract NAS 5–26555. These observations are associated with program HST-GO-15916 (PI: L. Matrà). The National Radio Astronomy Observatory is a facility of the National Science Foundation operated under a cooperative agreement by Associated Universities, Inc. Support for Program number HST-GO-15916.002-A was provided by NASA through a grant from the Space Telescope Science Institute, which is operated by the Association of Universities for Research in Astronomy, Incorporated, under NASA contract NAS5-26555.

\section{Data Availability}
The authors declare that the data supporting the findings of this study are available at \url{https://catalogs.mast.stsci.edu/hsc/}  under proposal ID 15916.

\bibliographystyle{mnras}
\bibliography{refs}

\begin{thebibliography}{}
\makeatletter
\relax
\def\mn@urlcharsother{\let\do\@makeother \do\$\do\&\do\#\do\^\do\_\do\%\do\~}
\def\mn@doi{\begingroup\mn@urlcharsother \@ifnextchar [ {\mn@doi@}
  {\mn@doi@[]}}
\def\mn@doi@[#1]#2{\def\@tempa{#1}\ifx\@tempa\@empty \href
  {http://dx.doi.org/#2} {doi:#2}\else \href {http://dx.doi.org/#2} {#1}\fi
  \endgroup}
\def\mn@eprint#1#2{\mn@eprint@#1:#2::\@nil}
\def\mn@eprint@arXiv#1{\href {http://arxiv.org/abs/#1} {{\tt arXiv:#1}}}
\def\mn@eprint@dblp#1{\href {http://dblp.uni-trier.de/rec/bibtex/#1.xml}
  {dblp:#1}}
\def\mn@eprint@#1:#2:#3:#4\@nil{\def\@tempa {#1}\def\@tempb {#2}\def\@tempc
  {#3}\ifx \@tempc \@empty \let \@tempc \@tempb \let \@tempb \@tempa \fi \ifx
  \@tempb \@empty \def\@tempb {arXiv}\fi \@ifundefined
  {mn@eprint@\@tempb}{\@tempb:\@tempc}{\expandafter \expandafter \csname
  mn@eprint@\@tempb\endcsname \expandafter{\@tempc}}}

\bibitem[\protect\citeauthoryear{{Beust}, {Lagrange-Henri}, {Vidal-Madjar}  \&
  {Ferlet}}{{Beust} et~al.}{1990}]{Beust(1990)}
{Beust} H.,  {Lagrange-Henri} A.~M.,  {Vidal-Madjar} A.,   {Ferlet} R.,  1990,
  \aap, \href {https://ui.adsabs.harvard.edu/abs/1990A&A...236..202B} {236,
  202}

\bibitem[\protect\citeauthoryear{{Bouret}, {Deleuil}, {Lanz}, {Roberge},
  {Lecavelier des Etangs}  \& {Vidal-Madjar}}{{Bouret}
  et~al.}{2002}]{Bouret(2002)}
{Bouret} J.~C.,  {Deleuil} M.,  {Lanz} T.,  {Roberge} A.,  {Lecavelier des
  Etangs} A.,   {Vidal-Madjar} A.,  2002, \mn@doi [\aap]
  {10.1051/0004-6361:20020741}, \href
  {https://ui.adsabs.harvard.edu/abs/2002A&A...390.1049B} {390, 1049}

\bibitem[\protect\citeauthoryear{{Cataldi} et~al.,}{{Cataldi}
  et~al.}{2020}]{Cataldi(2020)}
{Cataldi} G.,  et~al., 2020, \mn@doi [\apj] {10.3847/1538-4357/ab7cc7}, \href
  {https://ui.adsabs.harvard.edu/abs/2020ApJ...892...99C} {892, 99}

\bibitem[\protect\citeauthoryear{{Cataldi} et~al.,}{{Cataldi}
  et~al.}{2023}]{Cataldi(2023)}
{Cataldi} G.,  et~al., 2023, \mn@doi [arXiv e-prints]
  {10.48550/arXiv.2305.12093}, \href
  {https://ui.adsabs.harvard.edu/abs/2023arXiv230512093C} {p. arXiv:2305.12093}

\bibitem[\protect\citeauthoryear{{Crawford}, {Spyromilio}, {Barlow}, {Diego}
  \& {Lagrange}}{{Crawford} et~al.}{1994}]{Crawford(1994)}
{Crawford} I.~A.,  {Spyromilio} J.,  {Barlow} M.~J.,  {Diego} F.,   {Lagrange}
  A.~M.,  1994, \mn@doi [\mnras] {10.1093/mnras/266.1.L65}, \href
  {https://ui.adsabs.harvard.edu/abs/1994MNRAS.266L..65C} {266, L65}

\bibitem[\protect\citeauthoryear{{Deleuil} et~al.,}{{Deleuil}
  et~al.}{2001}]{Deleuil(2001)}
{Deleuil} M.,  et~al., 2001, \mn@doi [\apjl] {10.1086/323005}, \href
  {https://ui.adsabs.harvard.edu/abs/2001ApJ...557L..67D} {557, L67}

\bibitem[\protect\citeauthoryear{{Ferlet}, {Hobbs}  \& {Vidal-Madjar}}{{Ferlet}
  et~al.}{1987}]{Ferlet(1987)}
{Ferlet} R.,  {Hobbs} L.~M.,   {Vidal-Madjar} A.,  1987, \aap, \href
  {https://ui.adsabs.harvard.edu/abs/1987A&A...185..267F} {185, 267}

\bibitem[\protect\citeauthoryear{{Fern{\'a}ndez}, {Brandeker}  \&
  {Wu}}{{Fern{\'a}ndez} et~al.}{2006}]{Fernadez(2006)}
{Fern{\'a}ndez} R.,  {Brandeker} A.,   {Wu} Y.,  2006, \mn@doi [\apj]
  {10.1086/500788}, \href
  {https://ui.adsabs.harvard.edu/abs/2006ApJ...643..509F} {643, 509}

\bibitem[\protect\citeauthoryear{{Foreman-Mackey}, {Hogg}, {Lang}  \&
  {Goodman}}{{Foreman-Mackey} et~al.}{2013}]{Foreman-Mackey(2013)}
{Foreman-Mackey} D.,  {Hogg} D.~W.,  {Lang} D.,   {Goodman} J.,  2013, \mn@doi
  [\pasp] {10.1086/670067}, \href
  {https://ui.adsabs.harvard.edu/abs/2013PASP..125..306F} {125, 306}

\bibitem[\protect\citeauthoryear{{Gaia Collaboration} et~al.,}{{Gaia
  Collaboration} et~al.}{2016}]{Gaia(2016)}
{Gaia Collaboration} et~al., 2016, \mn@doi [\aap]
  {10.1051/0004-6361/201629512}, \href
  {https://ui.adsabs.harvard.edu/abs/2016A&A...595A...2G} {595, A2}

\bibitem[\protect\citeauthoryear{{Gaia Collaboration} et~al.,}{{Gaia
  Collaboration} et~al.}{2018}]{Gaia(2018)}
{Gaia Collaboration} et~al., 2018, \mn@doi [\aap]
  {10.1051/0004-6361/201833051}, \href
  {https://ui.adsabs.harvard.edu/abs/2018A&A...616A...1G} {616, A1}

\bibitem[\protect\citeauthoryear{{Goodman} \& {Weare}}{{Goodman} \&
  {Weare}}{2010}]{Goodman(2010)}
{Goodman} J.,  {Weare} J.,  2010, \mn@doi [Communications in Applied
  Mathematics and Computational Science] {10.2140/camcos.2010.5.65}, \href
  {https://ui.adsabs.harvard.edu/abs/2010CAMCS...5...65G} {5, 65}

\bibitem[\protect\citeauthoryear{{Hales}, {Barlow}, {Crawford}  \&
  {Casassus}}{{Hales} et~al.}{2017}]{A.Hales(2017)}
{Hales} A.~S.,  {Barlow} M.~J.,  {Crawford} I.~A.,   {Casassus} S.,  2017,
  \mn@doi [\mnras] {10.1093/mnras/stw3274}, \href
  {https://ui.adsabs.harvard.edu/abs/2017MNRAS.466.3582H} {466, 3582}

\bibitem[\protect\citeauthoryear{{Hales} et~al.,}{{Hales}
  et~al.}{2022}]{A.Hales(2022)}
{Hales} A.~S.,  et~al., 2022, \mn@doi [\apj] {10.3847/1538-4357/ac9cd3}, \href
  {https://ui.adsabs.harvard.edu/abs/2022ApJ...940..161H} {940, 161}

\bibitem[\protect\citeauthoryear{{Heays}, {Bosman}  \& {van Dishoeck}}{{Heays}
  et~al.}{2017}]{Heays(2017)}
{Heays} A.~N.,  {Bosman} A.~D.,   {van Dishoeck} E.~F.,  2017, \mn@doi [\aap]
  {10.1051/0004-6361/201628742}, \href
  {https://ui.adsabs.harvard.edu/abs/2017A&A...602A.105H} {602, A105}

\bibitem[\protect\citeauthoryear{{Henden} \& {Munari}}{{Henden} \&
  {Munari}}{2014}]{Henden(2014)}
{Henden} A.,  {Munari} U.,  2014, Contributions of the Astronomical Observatory
  Skalnate Pleso, \href {https://ui.adsabs.harvard.edu/abs/2014CoSka..43..518H}
  {43, 518}

\bibitem[\protect\citeauthoryear{{Higuchi} et~al.,}{{Higuchi}
  et~al.}{2017}]{Higuchi(2017)}
{Higuchi} A.~E.,  et~al., 2017, \mn@doi [\apjl] {10.3847/2041-8213/aa67f4},
  \href {https://ui.adsabs.harvard.edu/abs/2017ApJ...839L..14H} {839, L14}

\bibitem[\protect\citeauthoryear{{H{\o}g} et~al.,}{{H{\o}g}
  et~al.}{2000}]{Hog(2000)}
{H{\o}g} E.,  et~al., 2000, \aap, \href
  {https://ui.adsabs.harvard.edu/abs/2000A&A...355L..27H} {355, L27}

\bibitem[\protect\citeauthoryear{{Iglesias} et~al.,}{{Iglesias}
  et~al.}{2018}]{Iglesias(2018)}
{Iglesias} D.,  et~al., 2018, \mn@doi [\mnras] {10.1093/mnras/sty1724}, \href
  {https://ui.adsabs.harvard.edu/abs/2018MNRAS.480..488I} {480, 488}

\bibitem[\protect\citeauthoryear{{Kiefer}, {Lecavelier des Etangs}, {Boissier},
  {Vidal-Madjar}, {Beust}, {Lagrange}, {H{\'e}brard}  \& {Ferlet}}{{Kiefer}
  et~al.}{2014}]{Kiefer(2014)}
{Kiefer} F.,  {Lecavelier des Etangs} A.,  {Boissier} J.,  {Vidal-Madjar} A.,
  {Beust} H.,  {Lagrange} A.~M.,  {H{\'e}brard} G.,   {Ferlet} R.,  2014,
  \mn@doi [\nat] {10.1038/nature13849}, \href
  {https://ui.adsabs.harvard.edu/abs/2014Natur.514..462K} {514, 462}

\bibitem[\protect\citeauthoryear{{K{\'o}sp{\'a}l} et~al.,}{{K{\'o}sp{\'a}l}
  et~al.}{2013}]{Kospal(2013)}
{K{\'o}sp{\'a}l} {\'A}.,  et~al., 2013, \mn@doi [\apj]
  {10.1088/0004-637X/776/2/77}, \href
  {https://ui.adsabs.harvard.edu/abs/2013ApJ...776...77K} {776, 77}

\bibitem[\protect\citeauthoryear{{Kral} \& {Latter}}{{Kral} \&
  {Latter}}{2016}]{Kral&Latter(2016)}
{Kral} Q.,  {Latter} H.,  2016, \mn@doi [\mnras] {10.1093/mnras/stw1429}, \href
  {https://ui.adsabs.harvard.edu/abs/2016MNRAS.461.1614K} {461, 1614}

\bibitem[\protect\citeauthoryear{{Kral}, {Wyatt}, {Carswell}, {Pringle},
  {Matr{\`a}}  \& {Juh{\'a}sz}}{{Kral} et~al.}{2016}]{Q.Kral(2016)}
{Kral} Q.,  {Wyatt} M.,  {Carswell} R.~F.,  {Pringle} J.~E.,  {Matr{\`a}} L.,
  {Juh{\'a}sz} A.,  2016, \mn@doi [\mnras] {10.1093/mnras/stw1361}, \href
  {https://ui.adsabs.harvard.edu/abs/2016MNRAS.461..845K} {461, 845}

\bibitem[\protect\citeauthoryear{{Kral}, {Matr{\`a}}, {Wyatt}  \&
  {Kennedy}}{{Kral} et~al.}{2017}]{Q.Kral(2017)}
{Kral} Q.,  {Matr{\`a}} L.,  {Wyatt} M.~C.,   {Kennedy} G.~M.,  2017, \mn@doi
  [\mnras] {10.1093/mnras/stx730}, \href
  {https://ui.adsabs.harvard.edu/abs/2017MNRAS.469..521K} {469, 521}

\bibitem[\protect\citeauthoryear{{Kral}, {Marino}, {Wyatt}, {Kama}  \&
  {Matr{\`a}}}{{Kral} et~al.}{2019}]{Q.Kral(2019)}
{Kral} Q.,  {Marino} S.,  {Wyatt} M.~C.,  {Kama} M.,   {Matr{\`a}} L.,  2019,
  \mn@doi [\mnras] {10.1093/mnras/sty2923}, \href
  {https://ui.adsabs.harvard.edu/abs/2019MNRAS.489.3670K} {489, 3670}

\bibitem[\protect\citeauthoryear{{Kral}, {Pringle}, {Matr{\`a}}  \&
  {Th{\'e}bault}}{{Kral} et~al.}{2022}]{Kral(2022)}
{Kral} Q.,  {Pringle} J.,  {Matr{\`a}} L.,   {Th{\'e}bault} P.,  2022, \mn@doi
  [arXiv e-prints] {10.48550/arXiv.2211.04191}, \href
  {https://ui.adsabs.harvard.edu/abs/2022arXiv221104191K} {p. arXiv:2211.04191}

\bibitem[\protect\citeauthoryear{{Luhman} \& {Esplin}}{{Luhman} \&
  {Esplin}}{2020}]{Luhman(2020)}
{Luhman} K.~L.,  {Esplin} T.~L.,  2020, \mn@doi [\aj]
  {10.3847/1538-3881/ab9599}, \href
  {https://ui.adsabs.harvard.edu/abs/2020AJ....160...44L} {160, 44}

\bibitem[\protect\citeauthoryear{{MacGregor} et~al.,}{{MacGregor}
  et~al.}{2018}]{MacGregor(2018)}
{MacGregor} M.~A.,  et~al., 2018, \mn@doi [\apj] {10.3847/1538-4357/aaec71},
  \href {https://ui.adsabs.harvard.edu/abs/2018ApJ...869...75M} {869, 75}

\bibitem[\protect\citeauthoryear{{Mamajek} \& {Bell}}{{Mamajek} \&
  {Bell}}{2014}]{Mamajek(2014)}
{Mamajek} E.~E.,  {Bell} C. P.~M.,  2014, \mn@doi [\mnras]
  {10.1093/mnras/stu1894}, \href
  {https://ui.adsabs.harvard.edu/abs/2014MNRAS.445.2169M} {445, 2169}

\bibitem[\protect\citeauthoryear{{Marino} et~al.,}{{Marino}
  et~al.}{2016}]{Marino(2016)}
{Marino} S.,  et~al., 2016, \mn@doi [\mnras] {10.1093/mnras/stw1216}, \href
  {https://ui.adsabs.harvard.edu/abs/2016MNRAS.460.2933M} {460, 2933}

\bibitem[\protect\citeauthoryear{{Marino} et~al.,}{{Marino}
  et~al.}{2017}]{Marino(2017)}
{Marino} S.,  et~al., 2017, \mn@doi [\mnras] {10.1093/mnras/stw2867}, \href
  {https://ui.adsabs.harvard.edu/abs/2017MNRAS.465.2595M} {465, 2595}

\bibitem[\protect\citeauthoryear{{Marino}, {Flock}, {Henning}, {Kral},
  {Matr{\`a}}  \& {Wyatt}}{{Marino} et~al.}{2020}]{S.Marino(2020)}
{Marino} S.,  {Flock} M.,  {Henning} T.,  {Kral} Q.,  {Matr{\`a}} L.,   {Wyatt}
  M.~C.,  2020, \mn@doi [\mnras] {10.1093/mnras/stz3487}, \href
  {https://ui.adsabs.harvard.edu/abs/2020MNRAS.492.4409M} {492, 4409}

\bibitem[\protect\citeauthoryear{{Marino}, {Cataldi}, {Jankovic}, {Matr{\`a}}
  \& {Wyatt}}{{Marino} et~al.}{2022}]{S.Marino(2022)}
{Marino} S.,  {Cataldi} G.,  {Jankovic} M.~R.,  {Matr{\`a}} L.,   {Wyatt}
  M.~C.,  2022, \mn@doi [\mnras] {10.1093/mnras/stac1756}, \href
  {https://ui.adsabs.harvard.edu/abs/2022MNRAS.515..507M} {515, 507}

\bibitem[\protect\citeauthoryear{{Matr{\`a}}, {Pani{\'c}}, {Wyatt}  \&
  {Dent}}{{Matr{\`a}} et~al.}{2015}]{L.Matra(2015)}
{Matr{\`a}} L.,  {Pani{\'c}} O.,  {Wyatt} M.~C.,   {Dent} W.~R.~F.,  2015,
  \mn@doi [\mnras] {10.1093/mnras/stu2619}, \href
  {https://ui.adsabs.harvard.edu/abs/2015MNRAS.447.3936M} {447, 3936}

\bibitem[\protect\citeauthoryear{{Matr{\`a}} et~al.,}{{Matr{\`a}}
  et~al.}{2017}]{Matra(2017)}
{Matr{\`a}} L.,  et~al., 2017, \mn@doi [\apj] {10.3847/1538-4357/aa71b4}, \href
  {https://ui.adsabs.harvard.edu/abs/2017ApJ...842....9M} {842, 9}

\bibitem[\protect\citeauthoryear{{Matr{\`a}}, {Wilner}, {{\"O}berg}, {Andrews},
  {Loomis}, {Wyatt}  \& {Dent}}{{Matr{\`a}} et~al.}{2018}]{L.Matra(2018b)}
{Matr{\`a}} L.,  {Wilner} D.~J.,  {{\"O}berg} K.~I.,  {Andrews} S.~M.,
  {Loomis} R.~A.,  {Wyatt} M.~C.,   {Dent} W.~R.~F.,  2018, \mn@doi [\apj]
  {10.3847/1538-4357/aaa42a}, \href
  {https://ui.adsabs.harvard.edu/abs/2018ApJ...853..147M} {853, 147}

\bibitem[\protect\citeauthoryear{{Matr{\`a}}, {{\"O}berg}, {Wilner}, {Olofsson}
   \& {Bayo}}{{Matr{\`a}} et~al.}{2019}]{Matra(2019)}
{Matr{\`a}} L.,  {{\"O}berg} K.~I.,  {Wilner} D.~J.,  {Olofsson} J.,   {Bayo}
  A.,  2019, \mn@doi [\aj] {10.3847/1538-3881/aaff5b}, \href
  {https://ui.adsabs.harvard.edu/abs/2019AJ....157..117M} {157, 117}

\bibitem[\protect\citeauthoryear{{Matthews}, {Kennedy}, {Sibthorpe}, {Booth},
  {Wyatt}, {Broekhoven-Fiene}, {Macintosh}  \& {Marois}}{{Matthews}
  et~al.}{2014}]{Matthews(2014)}
{Matthews} B.,  {Kennedy} G.,  {Sibthorpe} B.,  {Booth} M.,  {Wyatt} M.,
  {Broekhoven-Fiene} H.,  {Macintosh} B.,   {Marois} C.,  2014, \mn@doi [\apj]
  {10.1088/0004-637X/780/1/97}, \href
  {https://ui.adsabs.harvard.edu/abs/2014ApJ...780...97M} {780, 97}

\bibitem[\protect\citeauthoryear{{Melis}, {Zuckerman}, {Rhee}, {Song}, {Murphy}
   \& {Bessell}}{{Melis} et~al.}{2013}]{C.Melis(2013)}
{Melis} C.,  {Zuckerman} B.,  {Rhee} J.~H.,  {Song} I.,  {Murphy} S.~J.,
  {Bessell} M.~S.,  2013, \mn@doi [\apj] {10.1088/0004-637X/778/1/12}, \href
  {https://ui.adsabs.harvard.edu/abs/2013ApJ...778...12M} {778, 12}

\bibitem[\protect\citeauthoryear{{Mo{\'o}r} et~al.,}{{Mo{\'o}r}
  et~al.}{2017}]{Moor(2017)}
{Mo{\'o}r} A.,  et~al., 2017, \mn@doi [\apj] {10.3847/1538-4357/aa8e4e}, \href
  {https://ui.adsabs.harvard.edu/abs/2017ApJ...849..123M} {849, 123}

\bibitem[\protect\citeauthoryear{{Morton} \& {Noreau}}{{Morton} \&
  {Noreau}}{1994}]{Morton(1994)}
{Morton} D.~C.,  {Noreau} L.,  1994, \mn@doi [\apjs] {10.1086/192100}, \href
  {https://ui.adsabs.harvard.edu/abs/1994ApJS...95..301M} {95, 301}

\bibitem[\protect\citeauthoryear{{Nakatani}, {Kobayashi}, {Kuiper}, {Nomura}
  \& {Aikawa}}{{Nakatani} et~al.}{2021}]{Nakatani(2021)}
{Nakatani} R.,  {Kobayashi} H.,  {Kuiper} R.,  {Nomura} H.,   {Aikawa} Y.,
  2021, \mn@doi [\apj] {10.3847/1538-4357/ac0137}, \href
  {https://ui.adsabs.harvard.edu/abs/2021ApJ...915...90N} {915, 90}

\bibitem[\protect\citeauthoryear{{Pecaut} \& {Mamajek}}{{Pecaut} \&
  {Mamajek}}{2016}]{Pecaut(2016)}
{Pecaut} M.~J.,  {Mamajek} E.~E.,  2016, \mn@doi [\mnras]
  {10.1093/mnras/stw1300}, \href
  {https://ui.adsabs.harvard.edu/abs/2016MNRAS.461..794P} {461, 794}

\bibitem[\protect\citeauthoryear{{Rebollido} et~al.,}{{Rebollido}
  et~al.}{2018a}]{rebollido(2018)}
{Rebollido} I.,  et~al., 2018a, \mn@doi [\aap] {10.1051/0004-6361/201732329},
  \href {https://ui.adsabs.harvard.edu/abs/2018A&A...614A...3R} {614, A3}

\bibitem[\protect\citeauthoryear{{Rebollido} et~al.,}{{Rebollido}
  et~al.}{2018b}]{I.Rebollido(2018)}
{Rebollido} I.,  et~al., 2018b, \mn@doi [\aap] {10.1051/0004-6361/201732329},
  \href {https://ui.adsabs.harvard.edu/abs/2018A&A...614A...3R} {614, A3}

\bibitem[\protect\citeauthoryear{{Redfield}}{{Redfield}}{2007}]{S.Redfield(2007)}
{Redfield} S.,  2007, \mn@doi [\apjl] {10.1086/512237}, \href
  {https://ui.adsabs.harvard.edu/abs/2007ApJ...656L..97R} {656, L97}

\bibitem[\protect\citeauthoryear{{Ricker} et~al.,}{{Ricker}
  et~al.}{2015}]{TESS(2015)}
{Ricker} G.~R.,  et~al., 2015, \mn@doi [Journal of Astronomical Telescopes,
  Instruments, and Systems] {10.1117/1.JATIS.1.1.014003}, \href
  {https://ui.adsabs.harvard.edu/abs/2015JATIS...1a4003R} {1, 014003}

\bibitem[\protect\citeauthoryear{{Roberge}, {Feldman}, {Lagrange},
  {Vidal-Madjar}, {Ferlet}, {Jolly}, {Lemaire}  \& {Rostas}}{{Roberge}
  et~al.}{2000}]{A.Roberge(2000)}
{Roberge} A.,  {Feldman} P.~D.,  {Lagrange} A.~M.,  {Vidal-Madjar} A.,
  {Ferlet} R.,  {Jolly} A.,  {Lemaire} J.~L.,   {Rostas} F.,  2000, \mn@doi
  [\apj] {10.1086/309157}, \href
  {https://ui.adsabs.harvard.edu/abs/2000ApJ...538..904R} {538, 904}

\bibitem[\protect\citeauthoryear{{Roberge}, {Feldman}, {Weinberger}, {Deleuil}
  \& {Bouret}}{{Roberge} et~al.}{2006}]{Roberge(2006)}
{Roberge} A.,  {Feldman} P.~D.,  {Weinberger} A.~J.,  {Deleuil} M.,   {Bouret}
  J.-C.,  2006, \mn@doi [\nat] {10.1038/nature04832}, \href
  {https://ui.adsabs.harvard.edu/abs/2006Natur.441..724R} {441, 724}

\bibitem[\protect\citeauthoryear{{Roberge}, {Welsh}, {Kamp}, {Weinberger}  \&
  {Grady}}{{Roberge} et~al.}{2014}]{A.Roberge(2014)}
{Roberge} A.,  {Welsh} B.~Y.,  {Kamp} I.,  {Weinberger} A.~J.,   {Grady} C.~A.,
   2014, \mn@doi [\apjl] {10.1088/2041-8205/796/1/L11}, \href
  {https://ui.adsabs.harvard.edu/abs/2014ApJ...796L..11R} {796, L11}

\bibitem[\protect\citeauthoryear{{Rostas} et~al.,}{{Rostas}
  et~al.}{2000}]{Rostas(2000)}
{Rostas} F.,  et~al., 2000, \mn@doi [Transactions of the International
  Astronomical Union, Series A] {10.1017/S0251107X00003345}, \href
  {https://ui.adsabs.harvard.edu/abs/2000IAUTA..24..380R} {24, 380}

\bibitem[\protect\citeauthoryear{{Sahu}}{{Sahu}}{1999}]{Sahu(1999)}
{Sahu} K.,  1999, in , Vol.~3, STIS Instrument Handbook for Cycle 9 v. 3.
p.~3

\bibitem[\protect\citeauthoryear{{Skrutskie} et~al.,}{{Skrutskie}
  et~al.}{2006}]{Skrutskie(2006)}
{Skrutskie} M.~F.,  et~al., 2006, \mn@doi [\aj] {10.1086/498708}, \href
  {https://ui.adsabs.harvard.edu/abs/2006AJ....131.1163S} {131, 1163}

\bibitem[\protect\citeauthoryear{{Smirnov-Pinchukov}, {Mo{\'o}r}, {Semenov},
  {{\'A}brah{\'a}m}, {Henning}, {K{\'o}sp{\'a}l}, {Hughes}  \& {di
  Folco}}{{Smirnov-Pinchukov} et~al.}{2022}]{Smirnov-Pinchukov(2022)}
{Smirnov-Pinchukov} G.~V.,  {Mo{\'o}r} A.,  {Semenov} D.~A.,  {{\'A}brah{\'a}m}
  P.,  {Henning} T.,  {K{\'o}sp{\'a}l} {\'A}.,  {Hughes} A.~M.,   {di Folco}
  E.,  2022, \mn@doi [\mnras] {10.1093/mnras/stab3146}, \href
  {https://ui.adsabs.harvard.edu/abs/2022MNRAS.510.1148S} {510, 1148}

\bibitem[\protect\citeauthoryear{{Troutman}, {Hinkle}, {Najita}, {Rettig}  \&
  {Brittain}}{{Troutman} et~al.}{2011}]{Troutman(2011)}
{Troutman} M.~R.,  {Hinkle} K.~H.,  {Najita} J.~R.,  {Rettig} T.~W.,
  {Brittain} S.~D.,  2011, \mn@doi [\apj] {10.1088/0004-637X/738/1/12}, \href
  {https://ui.adsabs.harvard.edu/abs/2011ApJ...738...12T} {738, 12}

\bibitem[\protect\citeauthoryear{{Vican}, {Schneider}, {Bryden}, {Melis},
  {Zuckerman}, {Rhee}  \& {Song}}{{Vican} et~al.}{2016}]{Vican(2016)}
{Vican} L.,  {Schneider} A.,  {Bryden} G.,  {Melis} C.,  {Zuckerman} B.,
  {Rhee} J.,   {Song} I.,  2016, \mn@doi [\apj] {10.3847/1538-4357/833/2/263},
  \href {https://ui.adsabs.harvard.edu/abs/2016ApJ...833..263V} {833, 263}

\bibitem[\protect\citeauthoryear{{Vines} \& {Jenkins}}{{Vines} \&
  {Jenkins}}{2022}]{Vines(2022)}
{Vines} J.~I.,  {Jenkins} J.~S.,  2022, \mn@doi [\mnras]
  {10.1093/mnras/stac956}, \href
  {https://ui.adsabs.harvard.edu/abs/2022MNRAS.tmp..920V} {}

\bibitem[\protect\citeauthoryear{{Worthen} et~al.,}{{Worthen}
  et~al.}{2023}]{K.Worthen(2023)}
{Worthen} K.,  et~al., 2023, \mn@doi [arXiv e-prints]
  {10.48550/arXiv.2312.09106}, \href
  {https://ui.adsabs.harvard.edu/abs/2023arXiv231209106W} {p. arXiv:2312.09106}

\bibitem[\protect\citeauthoryear{{Wright} et~al.,}{{Wright}
  et~al.}{2010}]{Wright(2010)}
{Wright} E.~L.,  et~al., 2010, \mn@doi [\aj] {10.1088/0004-6256/140/6/1868},
  \href {https://ui.adsabs.harvard.edu/abs/2010AJ....140.1868W} {140, 1868}

\bibitem[\protect\citeauthoryear{{Wyatt}, {Pani{\'c}}, {Kennedy}  \&
  {Matr{\`a}}}{{Wyatt} et~al.}{2015}]{Wyatt(2015)}
{Wyatt} M.~C.,  {Pani{\'c}} O.,  {Kennedy} G.~M.,   {Matr{\`a}} L.,  2015,
  \mn@doi [\apss] {10.1007/s10509-015-2315-6}, \href
  {https://ui.adsabs.harvard.edu/abs/2015Ap&SS.357..103W} {357, 103}

\bibitem[\protect\citeauthoryear{{Youngblood}, {Roberge}, {MacGregor},
  {Brandeker}, {Weinberger}, {P{\'e}rez}, {Grady}  \& {Welsh}}{{Youngblood}
  et~al.}{2021}]{Youngblood(2021)}
{Youngblood} A.,  {Roberge} A.,  {MacGregor} M.~A.,  {Brandeker} A.,
  {Weinberger} A.~J.,  {P{\'e}rez} S.,  {Grady} C.,   {Welsh} B.,  2021,
  \mn@doi [\aj] {10.3847/1538-3881/ac21d1}, \href
  {https://ui.adsabs.harvard.edu/abs/2021AJ....162..235Y} {162, 235}

\bibitem[\protect\citeauthoryear{{Zuckerman}}{{Zuckerman}}{2019}]{Zuckerman(2019)}
{Zuckerman} B.,  2019, \mn@doi [\apj] {10.3847/1538-4357/aaee66}, \href
  {https://ui.adsabs.harvard.edu/abs/2019ApJ...870...27Z} {870, 27}

\makeatother
\end{thebibliography}
\appendix
\section{Modelling}
For light travelling through a column of gas, in the absence of significant emission, the equation of radiative transfer can be solved to obtain
\begin{equation}
I_{\nu} (s)/I_{\nu} (0) = e^{-\tau_{ \nu }}
\end{equation}
where $I_{\nu} (0)$ (erg s$^{-1}$ cm$^{-2}$ \AA$^{-1}$) is the intensity of the background radiation (the host star in our case), $I_{\nu} (s)$ (erg s$^{-1} $cm$^{-2}$ \AA$^{-1}$) is the emerging radiation having crossed a column of gas in the edge-on disk along the line of sight to Earth, and $\tau_{ \nu }$ is the optical depth, which is the cross-section of a given transition, multiplied by the column density of gas in the lower level of the upward transition.  The cross-section is defined as
\begin{equation}
\sigma _{\nu } = \frac{h\nu _{ul}}{4\pi }B_{lu} V(\nu)
\end{equation}
where $h$ is Plank's constant (erg s), $\nu_{ul}$ (Hz) is the line rest frequency, $B_{lu}$ is the Einstein B coefficient, and $V(\nu)$ is the Voigt line profile\footnote{Implemented using \url{https://docs.scipy.org/doc/scipy/reference/generated/scipy.special.voigt_profile.html}}. The Voigt line profile is defined by
\begin{equation}
\gamma = \frac{A_{ul}}{4\pi},
\end{equation}
\begin{equation} 
\sigma =\frac{ \sqrt{log(2)} \Delta v}{\sqrt{2log(2)}}
\end{equation}
where $A_{ul}$ is the Einstein coefficient\footnote{For the CI absorption lines the Einstein coefficients are taken from the NIST database. For the CO absorption lines the Einstein coefficients are calculated.} and $\Delta \nu$ is the Doppler parameter.
The Einstein $B_{lu}$ coefficient is given by
\begin{equation}
B_{lu} = \frac{g_{u}}{g_{l}} \frac{A_{ul}c^{2}}{2h\nu_{0}^{3}}
\end{equation}
where $g_{u}$ is the degeneracy of the upper level, $g_{l}$ is the degeneracy of the lower level, and $c$ (cm s$^{-1}$) is the speed of light. For simplicity, we assumed that both the kinetic and excitation temperatures are constant throughout the entire column of gas. Additionally, we fit both the kinetic and excitation temperature as free parameters as the gas is not necessarily in LTE. The kinetic temperature contributes to the width of the line, where the Doppler parameter is given by
\begin{equation}
\Delta \nu = \frac{\nu_{0}}{c}\sqrt{\frac{2kT_{kin}}{m}}
\end{equation}
where $m$ (grams) is the mass of the gas species fitted, $k$ ({erg/K) is the Boltzmann's constant, and $T_{kin}$ (K) is the kinetic temperature. This Doppler parameter, along with natural and turbulence broadening, governs the line width, where the effect of turbulence and natural broadening are small. The energy levels are populated according to the Boltzmann distribution, leading to a fractional population given by
\begin{equation}
x_l =  \frac{N_{l}}{N_{tot}} =\frac{g_{l}}{Z}e^{-E_{l}/kT_{exc}}
\end{equation}
where $E_{l}$ (K) is the lower energy level, $T_{exc}$ (K) is the excitation temperature, and $Z$ is the partition function. The partition function indicates the average number of states that are accessible to a species at the excitation temperature of the system and is defined as
\begin{equation}
Z = \sum g_{i}e^{-E_{i}/kT_{exc}}
\end{equation}
Where the sum is over all energy levels $E_{i}$ considered.

\section{CO bands}

\begin{figure*}
  \includegraphics[width=1\linewidth]{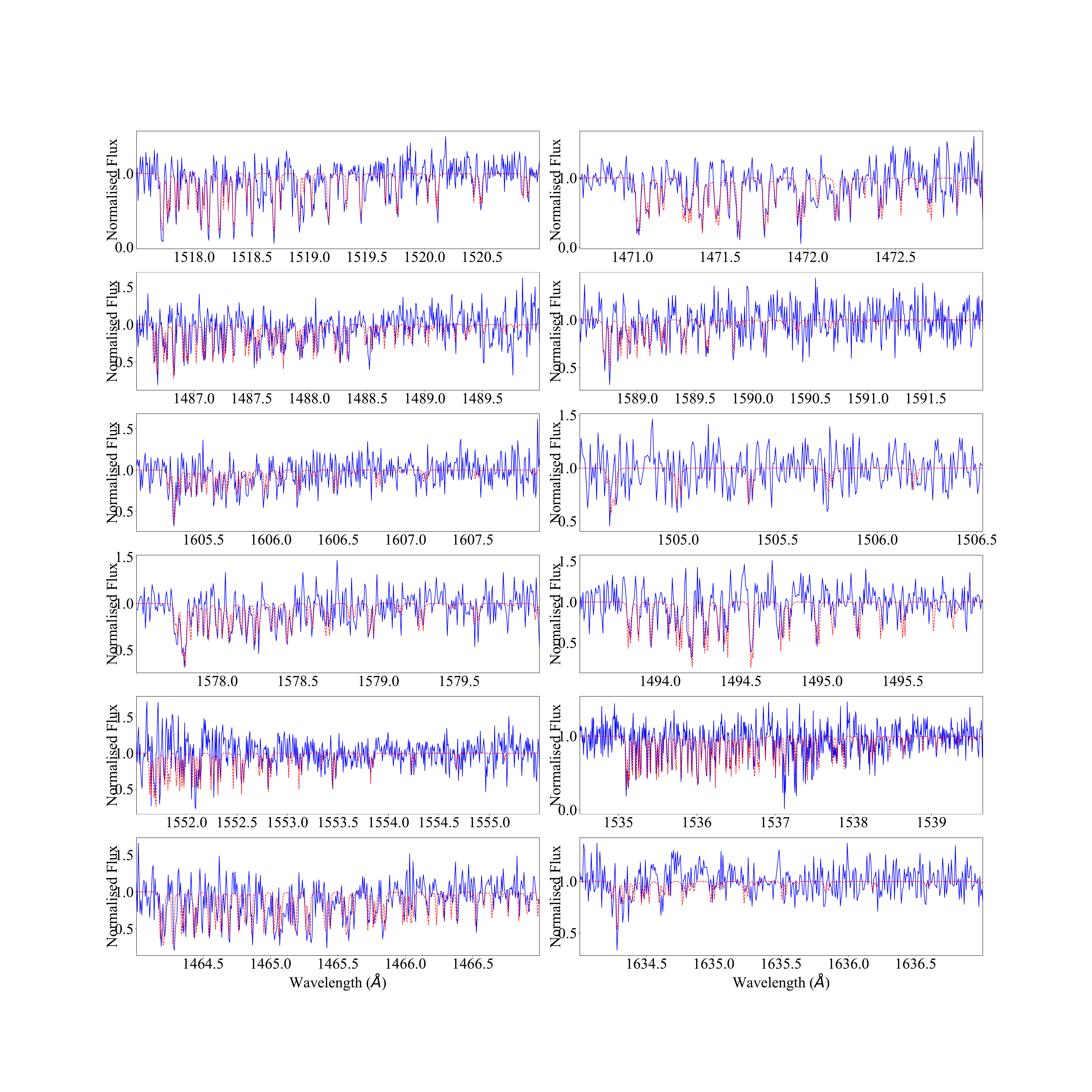}
  \caption{HST continuum-normalised data (blue) for different CO bands (Table \ref{table:lines}) overplotted with the best-fit models. From top left to bottom right: 1517.67 \AA, 1470.97 \AA, 1486.60 \AA, 1588.64 \AA, 1605.17 \AA, 1503.25 \AA, 1577.67 \AA, 1493.76 \AA,  1551.62 \AA, 1535.09 \AA, 1464.12 \AA, 1634.19 \AA.}
  \label{fig:HD110_CO}
\end{figure*}
\begin{figure*}
 \includegraphics[width=1\linewidth]{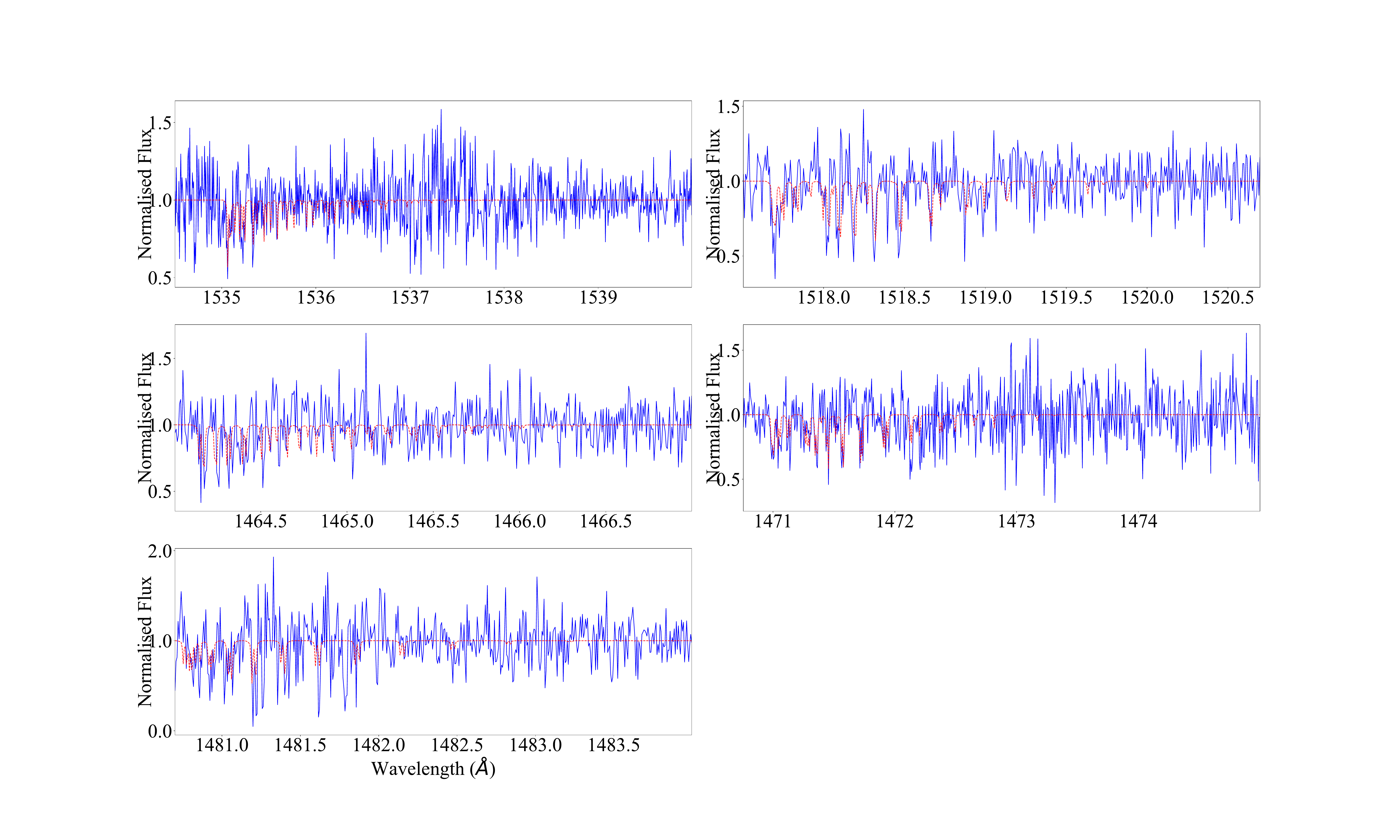}
 \caption{HST continuum-normalised data (blue) for different CO bands (Table \ref{table:lines}) overplotted with the best-fit models. From top left to bottom right: 1535.05 \AA, 1517.67 \AA, 1464.12 \AA, 1470.97 \AA, 1480.81 \AA.}
  \label{fig:HD131_CO}
\end{figure*}

\section{Corner plots}
\begin{figure*}
 \includegraphics[width=.8\linewidth]{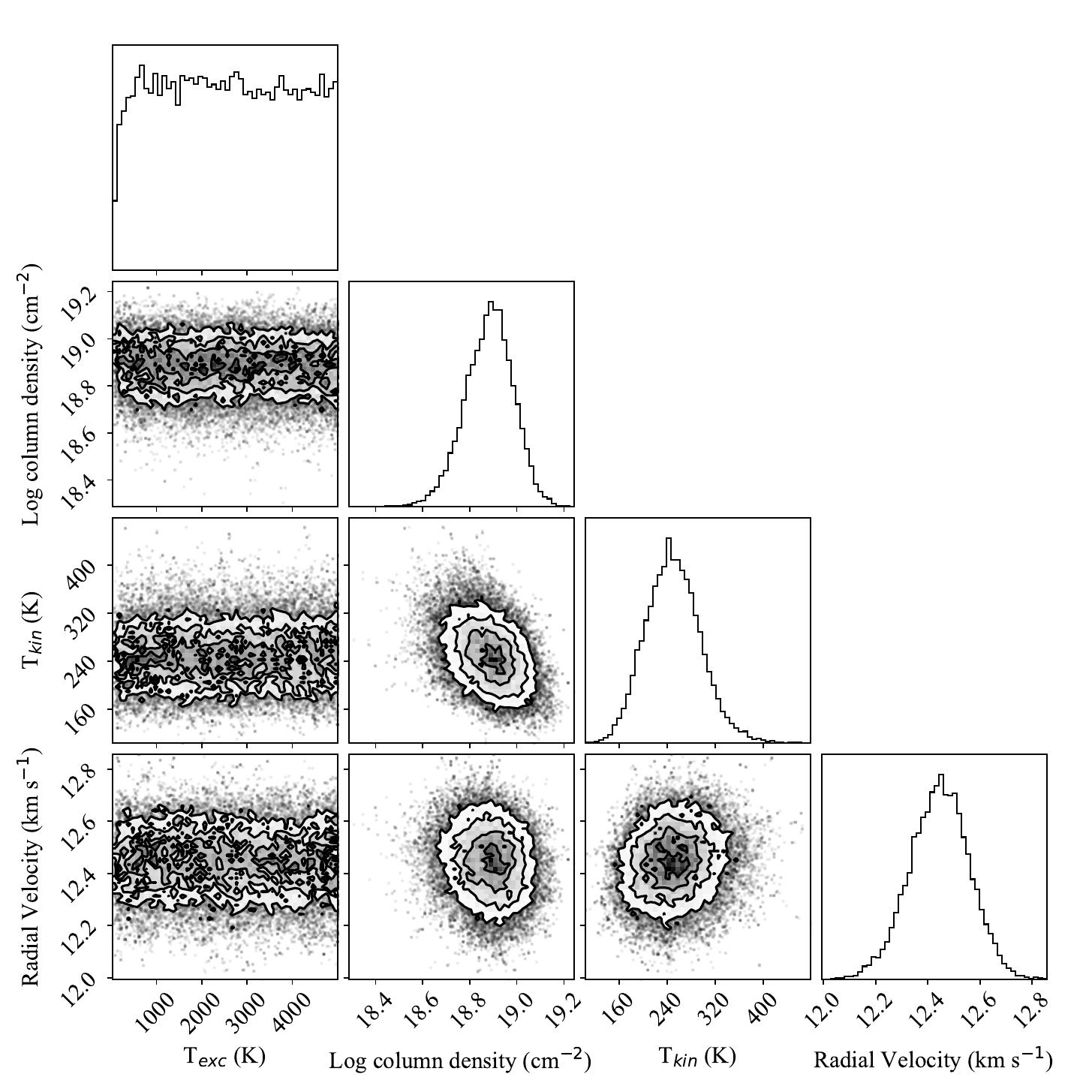}
 \caption{Posterior distribution of the simulated CI absorption model for HD110058, where total column density, kinetic temperature, excitation temperature and radial velocity are parameters. The marginalised distributions are presented in the diagonal. The best-fit values and uncertainties for the parameters are taken from the 16th, 50th and 84th percentiles.}
  \label{fig:HD110_C_corner}
\end{figure*}
\begin{figure*}
 \includegraphics[width=.8\linewidth]{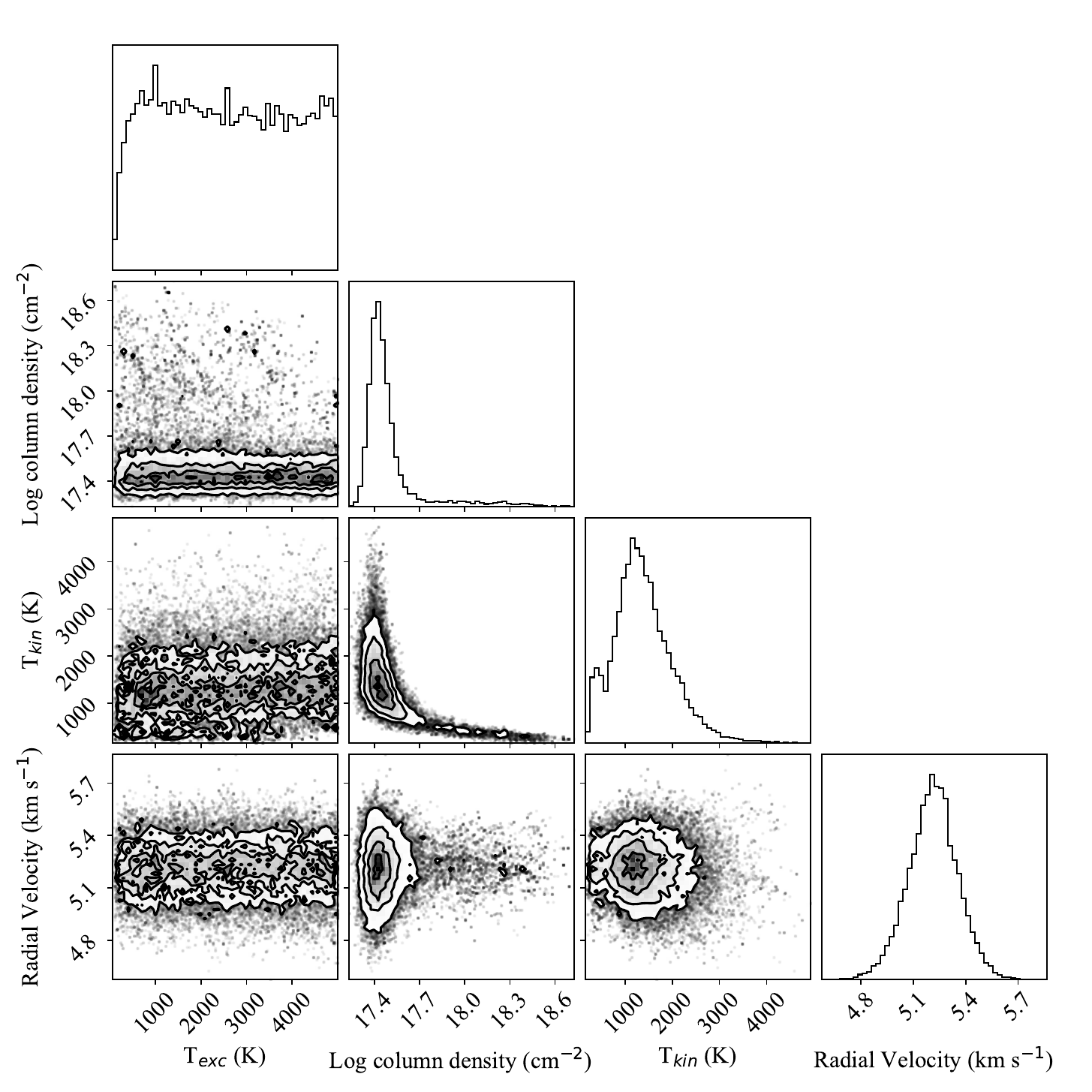}
 \caption{Posterior distribution of the simulated CI absorption model for HD131488, where total column density, kinetic temperature, excitation temperature and radial velocity are parameters. The marginalised distributions are presented in the diagonal. The best-fit values and uncertainties for the parameters are taken from the 16th, 50th and 84th percentiles.}
\label{fig:HD131_C_corner}
\end{figure*}
\begin{figure*}
 \includegraphics[width=.8\linewidth]{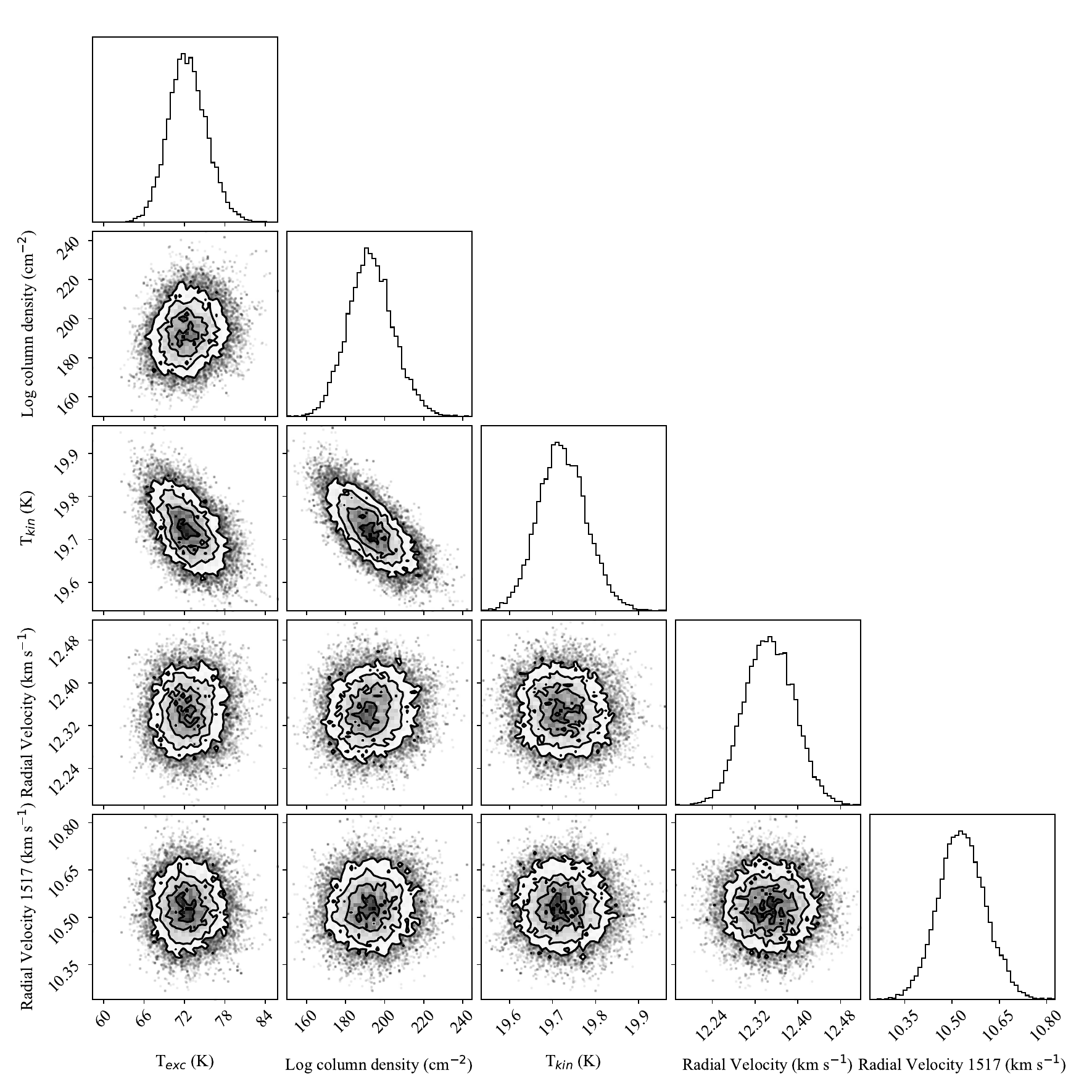}
 \caption{Posterior distribution of the simulated CO absorption model for HD110058, where total column density, kinetic temperature, excitation temperature and radial velocity are parameters. The radial velocity of one band (1517.67 \AA) was fitted separately due to a systematic offset, likely due to the wavelength calibration for this order (bands in all other orders are not affected). The marginalised distributions are presented in the diagonal. The best-fit values and uncertainties for the parameters are taken from the 16th, 50th and 84th percentiles.}
\label{fig:HD110_CO_corner}
\end{figure*}
\begin{figure*}
 \includegraphics[width=.8\linewidth]{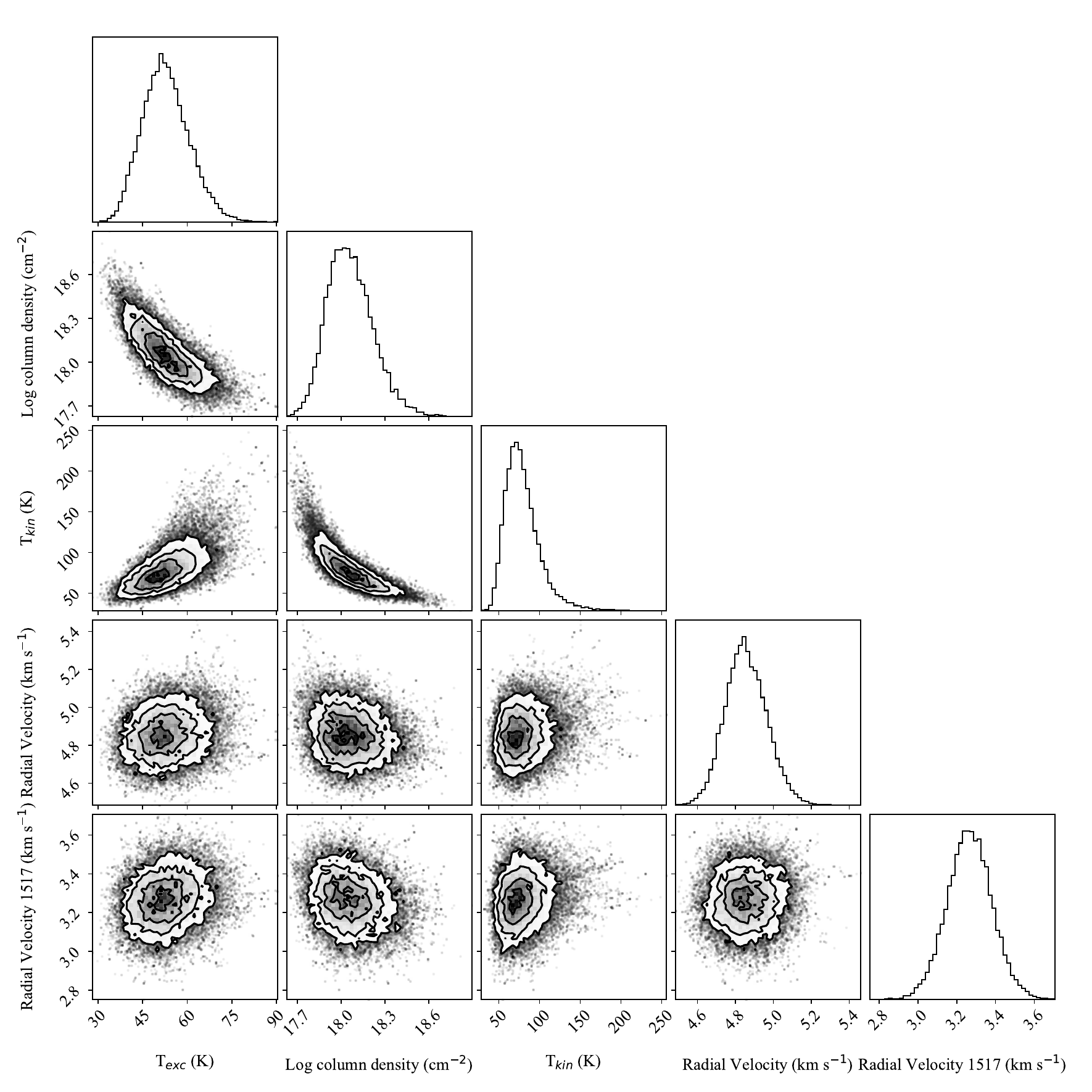}
 \caption{Posterior distribution of the simulated CO absorption model for HD131488, where total column density, kinetic temperature, excitation temperature and radial velocity are parameters. The radial velocity of one band (1517.67 \AA) was fitted separately (see comment in Fig. \ref{fig:HD110_CO_corner}). The marginalised distributions are presented in the diagonal. The best-fit values and uncertainties for the parameters are taken from the 16th, 50th and 84th percentiles.}
\label{fig:HD131_CO_corner}
\end{figure*}

\section{Stellar Fits}

\begin{figure*}
  \includegraphics[width=.6\linewidth]{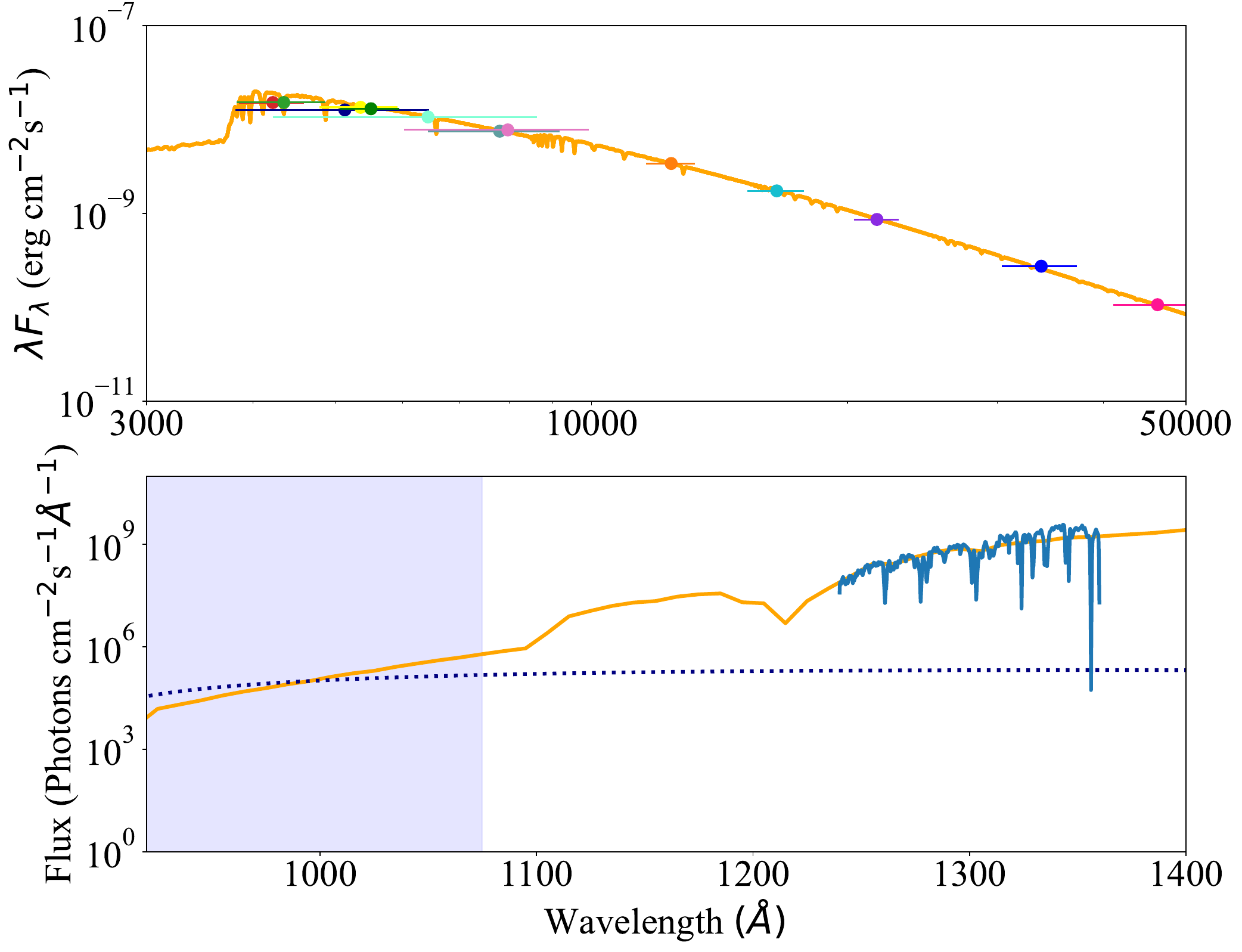}
  \caption{(Top) The best-fit model (orange line) for HD110058, as determined from fitting optical and near-IR photometry which are summarized in Table \ref{table:photometry_data} and plotted over the model. (Bottom) The best-fit model (orange line) rescaled to the COS data (blue solid line). Both the model and COS data have been rescaled to a distance of 22 au from the star, to compare with the ISRF (navy dotted line). The blue-shaded region corresponds to the wavelength range where photodissociation occurs.}
  \label{fig:stellar_HD110}
\end{figure*}
\begin{figure*}
  \includegraphics[width=.6\linewidth]{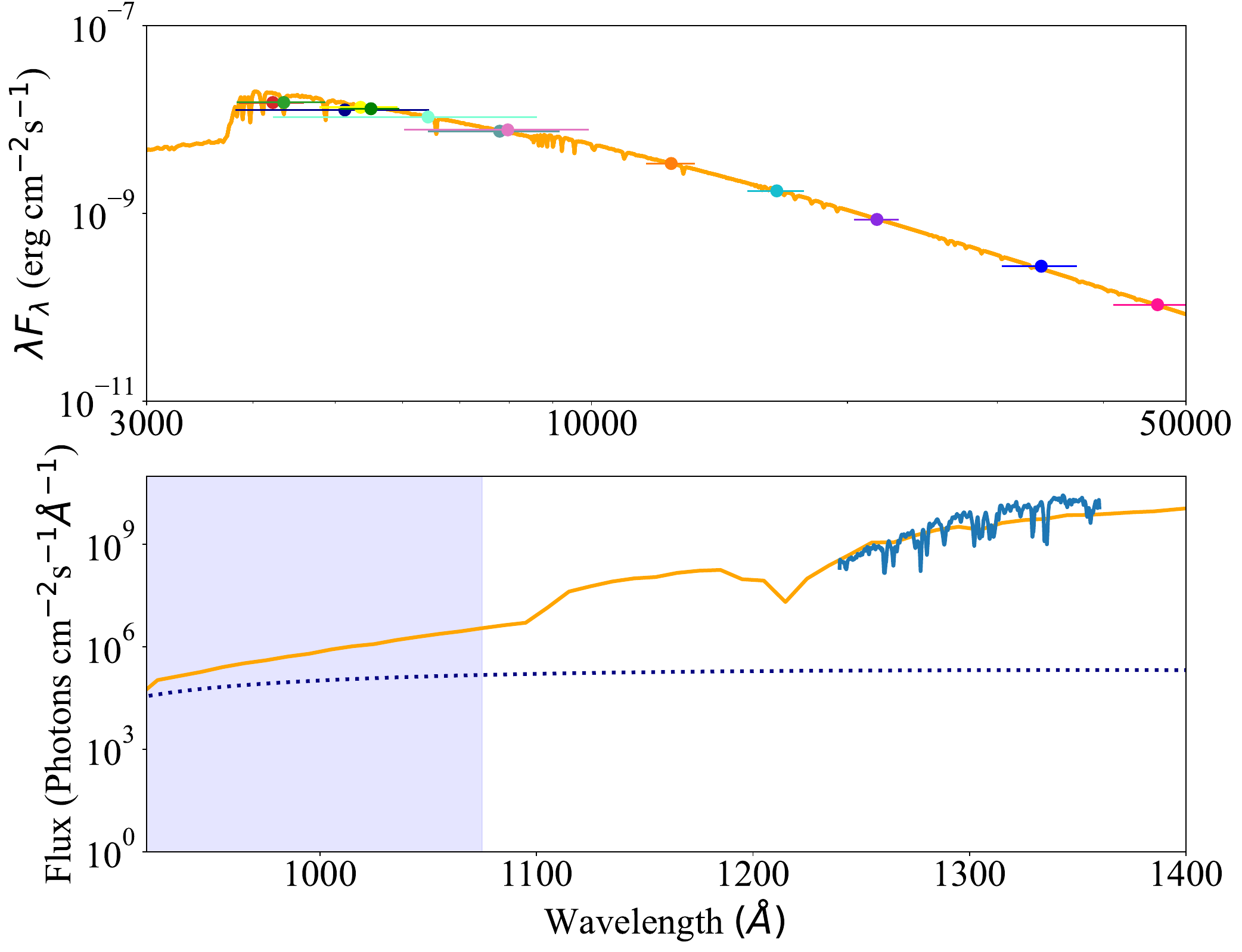}
  \caption{(Top) The best-fit model (orange line) for HD131488, as determined from fitting optical and near-IR photometry which are summarized in Table \ref{table:photometry_data}. (Bottom) The best-fit model (orange line) rescaled to the COS data (blue solid line). Both the model and COS data have been rescaled to a distance of 35 au from the star, to compare with the ISRF (navy dotted line). The blue-shaded region corresponds to the wavelength range where photodissociation occurs.}
  \label{fig:stellar_HD131}
\end{figure*}

\begin{table*}
\caption{Table of data points fitted in section 3.3.2 for both stars.}
\label{table:photometry_data}
\begin{tabular}{|ccccc|}
\hline
\multicolumn{1}{|l|}{Wavelength (\AA)} & \multicolumn{1}{c|}{Filter} & \multicolumn{1}{l|}{Magnitude (HD131488)} & \multicolumn{1}{l|}{Magnitude (HD110058)} & \multicolumn{1}{l|}{Reference} \\ \hline
\multicolumn{1}{|c|}{4220} & \multicolumn{1}{c|}{TYCHO B MvB} & \multicolumn{1}{c|}{8.12$\pm$0.016} & \multicolumn{1}{c|}{8.16$\pm$0.016} & (1) \\ \hline
\multicolumn{1}{|c|}{4350} & \multicolumn{1}{l|}{GROUND JOHNSON B} & \multicolumn{1}{c|}{8.10$\pm$0.01} & \multicolumn{1}{c|}{8.135$\pm$0.008} & (2) \\ \hline
\multicolumn{1}{|c|}{5130} & \multicolumn{1}{c|}{GaiaDR2v2 BP8.0515} & \multicolumn{1}{c|}{8.0515$\pm$0.0026} & \multicolumn{1}{c|}{8.039$\pm$0.0013} & (3) \\ \hline
\multicolumn{1}{|c|}{5350} & \multicolumn{1}{c|}{TYCHO V MvB} & \multicolumn{1}{c|}{8.012$\pm$0.013} & \multicolumn{1}{c|}{7.99$\pm$0.011} & (4) \\ \hline
\multicolumn{1}{|c|}{5500} & \multicolumn{1}{l|}{GROUND JOHNSON V} & \multicolumn{1}{c|}{8.003$\pm$0.013} & \multicolumn{1}{c|}{7.971$\pm$0.008} & (5) \\ \hline
\multicolumn{1}{|c|}{6420} & \multicolumn{1}{c|}{GaiaDR2v2 G} & \multicolumn{1}{c|}{7.9903$\pm$0.0007} & \multicolumn{1}{c|}{7.9500$\pm$0.0005} & (6) \\ \hline
\multicolumn{1}{|c|}{7800} & \multicolumn{1}{c|}{GaiaDR2v2 RP} & \multicolumn{1}{c|}{7.9367$\pm$0.0096} & \multicolumn{1}{c|}{7.8252$\pm$0.0016} & (7) \\ \hline
\multicolumn{1}{|c|}{7970} & \multicolumn{1}{c|}{TESS} & \multicolumn{1}{c|}{7.955$\pm$0.006} & \multicolumn{1}{c|}{7.854$\pm$0.006} & (8) \\ \hline
\multicolumn{1}{|c|}{12400} & \multicolumn{1}{c|}{2MASS J} & \multicolumn{1}{c|}{7.847$\pm$0.034} & \multicolumn{1}{c|}{7.643$\pm$0.023} & (9) \\ \hline
\multicolumn{1}{|c|}{16500} & \multicolumn{1}{c|}{2MASS H} & \multicolumn{1}{c|}{7.847$\pm$0.047} & \multicolumn{1}{c|}{7.587$\pm$0.033} & (10) \\ \hline
\multicolumn{1}{|c|}{21700} & \multicolumn{1}{c|}{2MASS Ks} & \multicolumn{1}{c|}{7.803$\pm$0.033} & \multicolumn{1}{c|}{7.580$\pm$0.018} & (11) \\ \hline
\multicolumn{1}{|c|}{33800} & \multicolumn{1}{c|}{WISE RSR W1} & \multicolumn{1}{c|}{7.717$\pm$0.029} & \multicolumn{1}{c|}{7.523$\pm$0.031} & (12) \\ \hline
\multicolumn{1}{|c|}{46300} & \multicolumn{1}{c|}{WISE RSR W2} & \multicolumn{1}{c|}{7.56$\pm$0.02} & \multicolumn{1}{c|}{7.56$\pm$0.02} & (13) \\ \hline
\multicolumn{5}{|l|}{\begin{tabular}[c]{@{}l@{}}(1)(4)\cite{Hog(2000)} (2)(5)\cite{Henden(2014)} (3)(6)(7)\cite{Gaia(2016)} \\
\cite{Gaia(2018)} (8)\cite{TESS(2015)}, (9)(10)(11)\cite{Skrutskie(2006)},
(12)(13)\cite{Wright(2010)}  \end{tabular}}\\ \hline
\end{tabular}
\end{table*}

\bsp	
\label{lastpage}
\end{document}